\documentclass[a4paper]{scrartcl}
\usepackage{upgreek}
\usepackage{amsmath}
\usepackage{amsfonts}
\usepackage{amstext}
\usepackage{mathtools}
\usepackage{amssymb}
\usepackage{siunitx}
\usepackage{float}
\usepackage{subcaption}
\usepackage{rotating}
\usepackage{setspace}
\usepackage{xcolor}
\usepackage{mathrsfs}
\usepackage{upgreek}
\usepackage{graphicx}
\usepackage[square, comma, numbers, sort&compress]{natbib} 
\usepackage{authblk}
\usepackage{hyperref}
\usepackage{lineno}
\newcommand{\autocite}[1]{\cite{#1}}
\begin{document}
    \title{The onset of molecule-spanning dynamics in a multi-domain protein}
	\author[1]{Benedikt Sohmen}
	\author[2,3]{Christian Beck}
	\author[3]{Tilo Seydel}
	\author[3]{Ingo Hoffmann}
	\author[1]{Bianca Hermann}
	\author[4]{Mark N\"uesch}
	\author[3]{Marco Grimaldo}
	\author[2]{Frank Schreiber$^*$}
	\author[5]{Steffen Wolf$^*$}
	\author[6]{Felix Roosen-Runge$^*$}
	\author[1,7]{Thorsten Hugel$^*$}
	\affil[1]{\normalsize Institute of Physical Chemistry, University of Freiburg, Albertstrasse 21, 79104 Freiburg, Germany}
	\affil[2]{Institute of Applied Physics, University of T\"ubingen,  Auf der Morgenstelle 10, 72076 T\"ubingen, Germany}
	\affil[3]{Institut Max von Laue - Paul Langevin, 71 avenue des Martyrs, 38042 Grenoble, France}
	\affil[4]{Department of Biochemistry, University of Zurich, Winterthurerstrasse 190, CH-8057 Zurich, Switzerland}
	\affil[5]{Biomolecular Dynamics, Institute of Physics, University of Freiburg, Hermann-Herder-Strasse 3, 79104 Freiburg, Germany}
	\affil[6]{Department of Biomedical Sciences and Biofilms-Research Center for Biointerfaces (BRCB), Malm\"o University, 20506 Malm\"o, Sweden}
	\affil[7]{Signalling Research Centers BIOSS and CIBSS, University of Freiburg, Sch\"anzlestrasse 18, 79104 Freiburg, Germany}
	\affil[ ]{$^*$contact details:
		\href{mailto:frank.schreiber@uni-tuebingen.de}{frank.schreiber@uni-tuebingen.de};
		 \href{mailto:steffen.wolf@physik.uni-freiburg.de}{steffen.wolf@physik.uni-freiburg.de};
		 \href{mailto:felix.roosen-runge@mau.se}{felix.roosen-runge@mau.se};
		 \href{mailto:thorsten.hugel@physchem.uni-freiburg.de}{thorsten.hugel@physchem.uni-freiburg.de}}
	\maketitle

\section*{Abstract}
	Protein dynamics has been investigated on a wide range of time scales. Nano- and picosecond dynamics have been assigned to local fluctuations, while slower dynamics have been attributed to larger conformational changes. However, it is largely unknown how local fluctuations can lead to global allosteric changes. Here we show that molecule-spanning dynamics on the 100\,ns time scale precede larger allosteric changes. We assign global real-space movements to  dynamic modes on the 100\,ns time scales, which became possible by a combination of single-molecule fluorescence, quasi-elastic neutron scattering and all-atom MD simulations. Additionally, we demonstrate the effect of Sba1, a co-chaperone of Hsp90, on these molecule-spanning dynamics, which implies functional importance of such dynamics. Our integrative approach provides comprehensive insights into molecule-spanning dynamics on the nanosecond time scale for a multi-domain protein and indicates that such dynamics are the molecular basis for allostery and large conformational changes in proteins.	

\section*{Main}

The complexity of understanding protein function results from the involvement of dynamic processes occurring on a broad range of time scales \autocite{Kern.2021}. Numerous studies have provided important insights into dynamics on the $\upmu$s to sec time scale. Protein binding kinetics and transition between defined conformational states have been successfully investigated\autocite{Callaway.2013, Ratzke.2014, Meli.2016, Barth.2018, Sanabria.2020, Schulze.2016}.  However, it becomes more and more clear that the dynamics within conformational states on time scales of a few to several hundreds of nanoseconds are also crucial as these might eventually enable and drive conformational transitions\autocite{Bozovic.2020, Bozovic.2021, Hu.2015}. While local dynamics on the picosecond time scale are well investigated for Hsp90 \autocite{Lopez.2021} global conformational changes could not be accessed and linked to spatial information on nanosecond time scales.

\begin{figure}[ht]
	\includegraphics[width=0.95\textwidth]{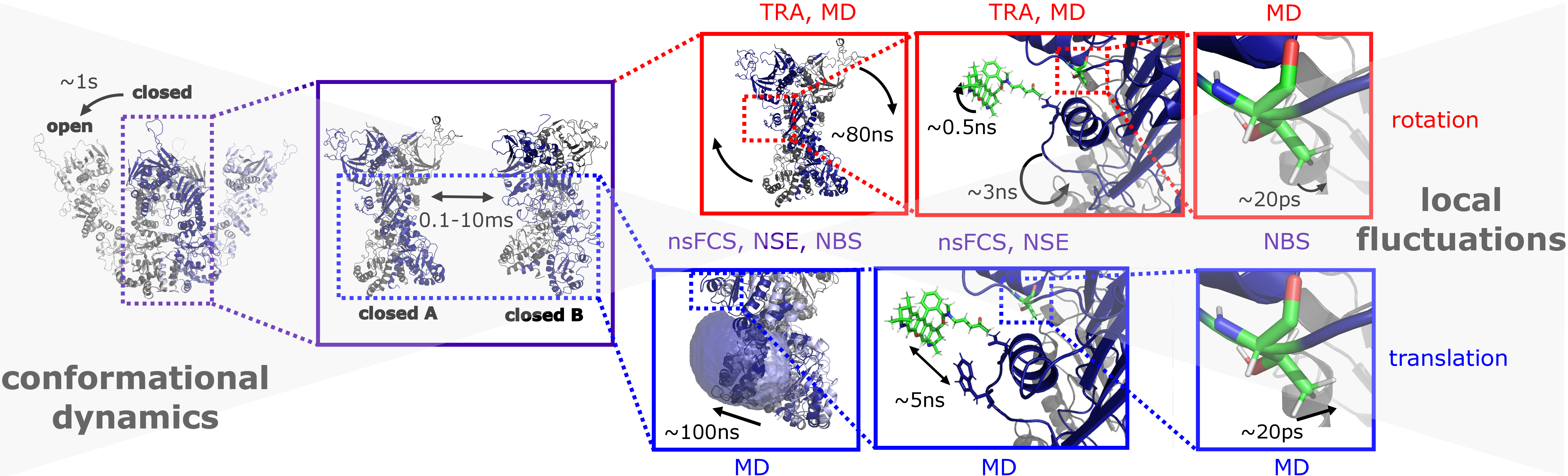}
	\caption{Comprehensive dynamic picture of Hsp90. Here, we complete the dynamical picture of Hsp90 by accessing the 100\,ns time scale which we will show to unravel molecule-spanning dynamics. Disentanglement of rotational and translational dynamics is feasible by combining nanosecond fluorescence correlation spectroscopy (nsFCS), time-resolved anisotropy (TRA), neutron spin echo spectroscopy (NSE), neutron backscattering spectroscopy (NBS) and molecular dynamic (MD) simulations. The time scales of the respective dynamics are given in the boxes.}
	\label{fig:timescale-overview}
\end{figure}

Here, we address this fundamental and highly relevant challenge using the well-established multi-domain protein Hsp90 (see Fig.\,\ref{fig:timescale-overview}). As a molecular chaperone Hsp90 helps proteins to find their native, biologically active conformation and mediates signal transduction\autocite{Taipale.2010, Schopf.2017}. Due to its interaction with several co-chaperones and several 100 client proteins, Hsp90 is considered a promising drug target to fight diseases such as cancer, Alzheimer and diabetes\autocite{Neckers.2012}.

Yeast Hsp90 is a 82\,kDa homodimer with each monomer consisting of 670 amino acids and three domains: The C-domain, responsible for dimerisation, the M-domain, hosting important co-chaperone and client binding sites and the N-domain, harbouring a nucleotide binding site and co-chaperone and client binding sites\autocite{Girstmair.2019}. Hsp90 is known to cycle through open and closed conformations on time scales ranging at least from milliseconds to many minutes, which manifests its high functionality\autocite{Shiau.2006, Graf.2009, Mickler.2009, Seifert.2012, Schopf.2017, Ye.2018}. Despite its importance, the mechanistic origin of this function-related molecular flexibility has not been understood, and insights into intra-state protein dynamics are thus promising to better understand the molecular basis for the transitions between conformations.

Given the complexity to reliably characterize nanosecond protein dynamics of Hsp90, we use an integrative approach combining several experimental techniques and computer simulations (outlined in Fig.\,S1).
Fluorescence correlation spectroscopy (FCS) probes correlated dynamics based on characteristic decays or rises of the fluorescence intensity correlation function \autocite{Rigler.1993, Felekyan.2005}. In the last decade, the potential of FCS on the nanosecond time scale was successfully demonstrated by studying unfolded proteins \autocite{Borgia.2012, Haenni.2013, Zosel.2017, Krainer.2017, Nueesch.2022}. Important pre-requisites for clear data interpretation are the separation of time scales \autocite{Gopich.2009} and the complementary distance information from FRET. Labeling with FRET donors and acceptors provided additional information by cross-correlating the donor and acceptor signal:
Signals could clearly be assigned to nanometer-scale structural changes if the donor-acceptor cross-correlation was anti-correlated \autocite{Nettels.2008}. Significant anti-correlations were observed for small intrinsically disordered proteins (IDPs), but similar clear signatures from nsFCS have not yet been observed for multi-domain proteins. Despite the significant importance of multi-domain proteins as regulators, the usefulness of nsFCS for these proteins so far is still elusive.

Time-resolved anisotropy (TRA) probes rotational dynamics of dye-coupled bio\-mole\-cules by time-dependent and polarisation-sensitive detection of fluorescence intensities \autocite{Ha.1999}. Combining pulsed interleaved excitation (PIE) with time-correlated single photon counting (TCSPC) enables studying the rotation of labeled proteins as shown for small IDPs \autocite{Tsytlonok.2019, Sanabria.2020}.
In multi-domain proteins, the situation is slightly more complex because multiple rotational modes can interfere. The question arises how the hierarchical nature of rotational dynamics encodes protein function or even adds another level of protein regulation. In principle, time-resolved anisotropy contains all information necessary to answer this question, but the signals from independent effects are superimposed and need to be disentangled.
In the present work, we show that the determination of hierarchical rotational dynamics and solving these challenges is feasible with extraordinary statistics and a sophisticated fit model \autocite{Schroder.2005}.

Quasi-elastic neutron scattering (QENS) simultaneously accesses spatial and time correlations by probing the scattering function depending on the momentum $\hbar q$ and energy transfer $\hbar \omega$ of the neutron during the interaction with the sample. QENS constitutes a label-free, non-invasive and non-destructive technique which measures a ensemble-averaged signal with an unambiguous interpretation in terms of statistical mechanics. Most importantly, the QENS signal contains information on the diffusive motions, which, due to their dependence on $q$, can be directly associated with length scales of structural motions and spatial confinement. Based on the distinction between so-called coherent and incoherent scattering associated with the neutron and nuclear spin statistics involved in the scattering, both self- (tracer) and collective (mutual) diffusion processes can be quantified. Early work on proteins focused on dynamics in hydrated powders to target aspects of the water-protein coupling\autocite{Gabel.2002, Tarek.2002}.
Recent advances in neutron instrumentation allow to obtain signals from protein solutions that are sufficiently strong to separate the different superimposed contributions from localized as well as rotational and translational center-of-mass diffusive motions of proteins in aqueous solutions~\autocite{Grimaldo_2019_QuartRevBiophys}.
Importantly, with neutron spin echo (NSE) we access these molecule-spanning multi-domain motions unambiguously on well-defined nanosecond time and nanometer length scales\autocite{Biehl2011, Bu.2011, Stadler.2014, Hoffmann2014, Grimaldo_2019_QuartRevBiophys}.

We complement our data interpretation by full-atom molecular dynamic (MD) simulations, which are ideal to probe molecular details on these fast time scales, as time scales of hundreds of nanoseconds are readily accessible even for large proteins with sufficient sampling. We use MD simulations to investigate both local details such as the free volume at specific sites as well as molecule-spanning dynamic modes\autocite{Smith.2018}. The measured dynamics is neither purely Brownian dynamics nor can it be described by a simple harmonic oscillator. We access a complex pattern of diffusive modes, which can be described with our integrative approach covering a wide range of time and length scales. Overall, together with the previously published dynamics on slower time scales \autocite{Wolf.2021, Ye.2018, Schmid.2016} our results complete a comprehensive picture of Hsp90's dynamics from the nanosecond to second time scale and show that even the fastest dynamics is relevant for the molecular regulation mechanism controlled by an interacting protein.

\section*{Results and Discussion}

\begin{figure}[ht]
	\includegraphics[width=0.95\textwidth]{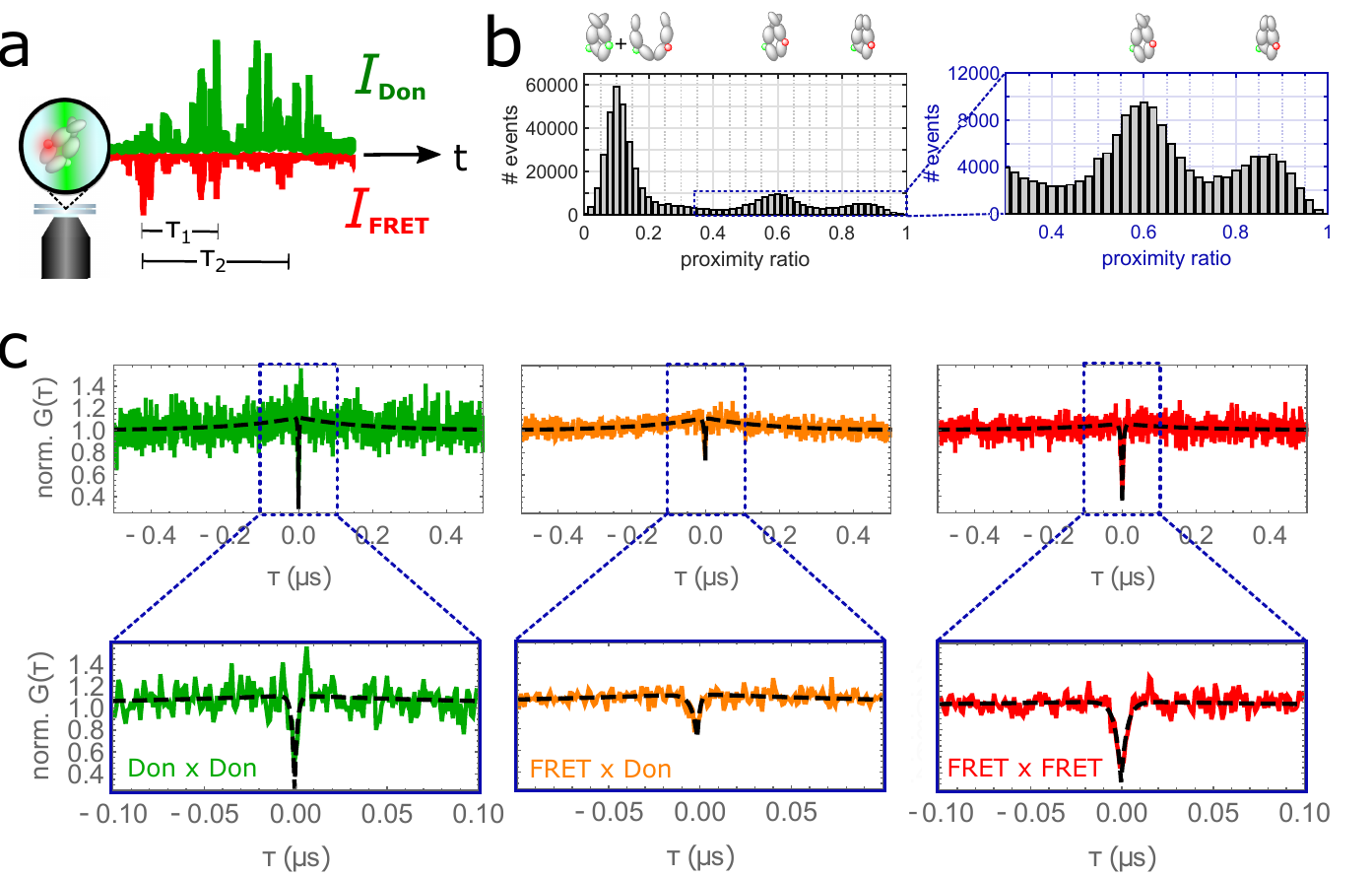}
	\caption{Substate-specific nsFCS analysis of the Hsp90 FRET pair 452-452 with AMPPNP. Atto532 and Atto647N were used as donor and acceptor molecules, respectively. a) Scheme of a nsFCS experiment. Donor- and acceptor-labeled proteins in the confocal volume are excited with green continuous wave (cw) illumination. Colour-sensitive detection discriminates donor-fluorescence ($I_{\textrm{Don}}$) and FRET-based acceptor fluorescence ($I_{\textrm{FRET}}$) with high time-resolution. The temporal correlation function $G(\tau)$ is calculated from the intensities. b) Substate-selection based on the proximity ratio $PR$. The $PR$-histogram reveals three populations at $\sim$0.1, $\sim$0.6 and $\sim$0.86. Selecting events with $PR>$0.3 enables specific analysis of the closed Hsp90 conformations. Note that at $PR\sim$0.1, the signal from donor-only molecules and FRET-labeled open Hsp90 molecules is superposed.  c) From a global analysis of all three channel correlations (Don$\times$Don, FRET$\times$Don and FRET$\times$FRET) we obtain a correlation time of 174$\pm$33\,ns for the closed conformations. Fluorophore antibunching occurs at 1.4$\pm$0.3\,ns and 3.1$\pm$0.4\,ns for donor and acceptor dyes, respectively. Errors were determined as standard fit errors. For the complete nsFCS analysis please refer to Fig.\,S5 and Tab.\,S1.}
	\label{fig:nsFCS_452-452}
\end{figure}

\subsection*{nsFCS reveals Hsp90 dynamics on the $\sim$100\,ns time scale}

nsFCS experiments were performed for three FRET pairs spanning the M-C domain of the Hsp90 dimer (298-298, 298-452 and 452-452). While FRET pairs with position 452 report on dynamics across the lumen of the Hsp90 dimer which is relevant for substrate and co-chaperone binding\autocite{Schopf.2017}, we chose position 298 because it is close to the functionally important `Hsp90 switch point' W300 which mediates client interactions\autocite{Rutz.2018}. Positions 298 and 452 can be labelled with fluorophores and are still functional as has been shown before\autocite{Hellenkamp.2017}.
All measurements were conducted in the presence of 2\,mM AMPPNP to populate the closed conformational states of Hsp90. Fig.\,\ref{fig:nsFCS_452-452} shows a representative nsFCS data set for the FRET pair 452-452 with AMPPNP. For a summary of all investigated Hsp90 FRET pairs see Fig.\,S5 and Tab. S1. Following the approach of separation of time scales\autocite{Gopich.2009} we obtain correlations on the $\sim$3\,ns and $\sim$100\,ns time scale. For experiments which involve position 298 a third correlation is detected at $\sim$5\,ns which we attribute to tryptophan quenching (see Fig. S6). The $\sim$3\,ns component is consistent with the manufacturer-specified fluorescence lifetime of the used dyes and given its characteristic feature it can be attributed to fluorophore antibunching. The $\sim$100\,ns component is present in both autocorrelations (Don$\times$Don and FRET$\times$FRET), as well as in the cross-correlation (Don$\times$FRET). Because auto- and cross-correlations do not show a clear anti-correlative behaviour a directly opposed structural movement between the investigated positions cannot explain this signal\autocite{Nettels.2008}. More likely, the correlation is caused by internal dynamics of Hsp90 subdomains, rotation of the overall protein or a superposition thereof. Such dynamics could for example affect the brightness of the donor (e.g. by quenching) and therefore show a correlation in all three signals.
For further clarification, we extended our experiments by measuring four different single-labeled Hsp90 variants with and without AMPPNP (Fig.\,S7). This allowed us to compare the correlation for Hsp90 in its closed and open conformation, respectively. Interestingly we find faster correlations in absence of AMPNP, when Hsp90 is mainly opened. Furthermore, we find that the correlation time depends on the dye location investigated. Altogether, our experiments disfavor global rotation as main cause of the correlation on the time scale of about $\sim$100\,ns and prompts us to propose a significant contribution from global internal Hsp90 dynamics, which could be confirmed by anisotropy decay experiments and MD simulations (see below).

\subsection*{Separating rotation from internal dynamics by time-resolved anisotropy}

\begin{figure}[ht]
	\includegraphics[width=0.95\textwidth]{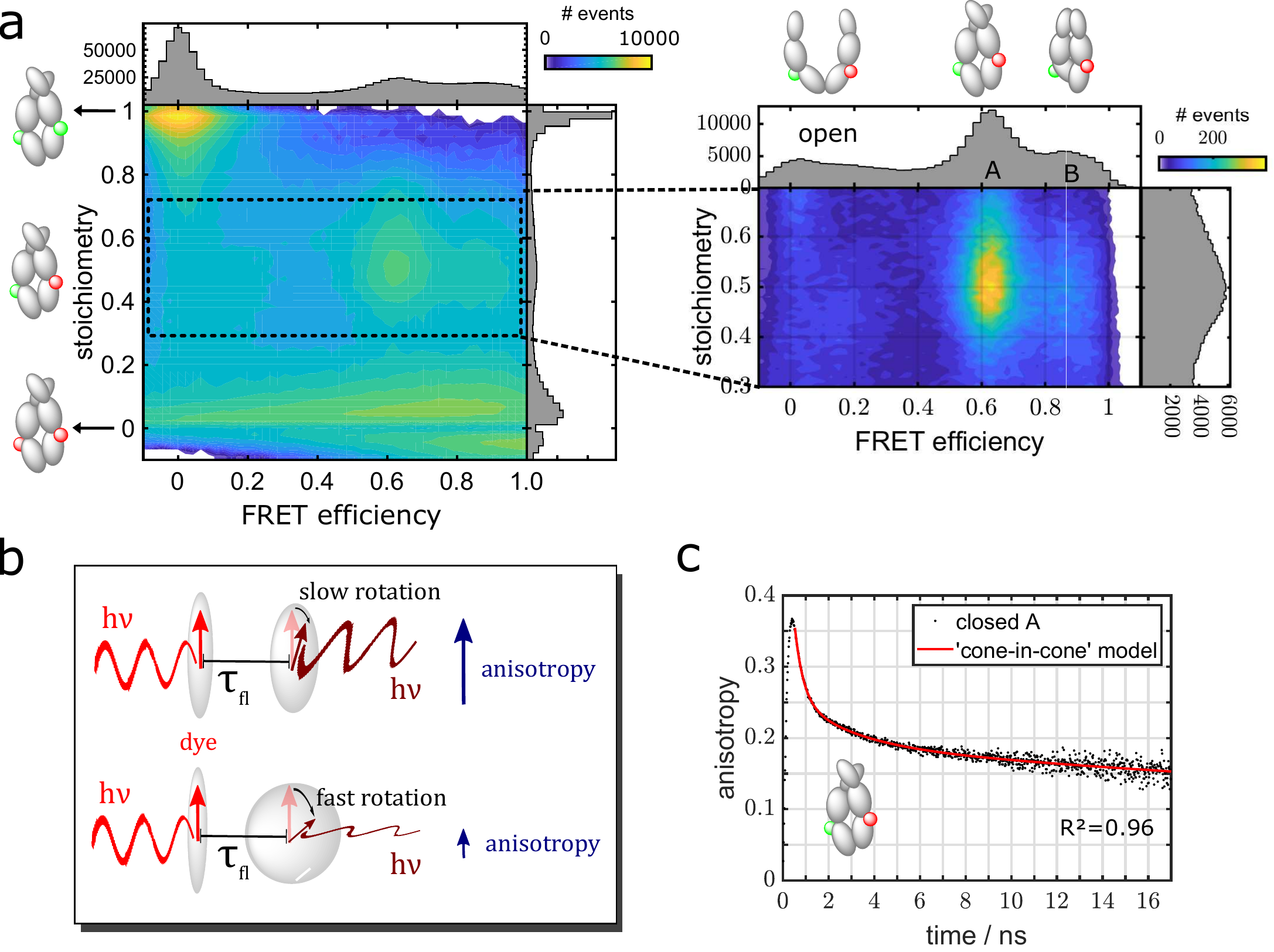}
	\caption{Substate-specific time-resolved single-molecule anisotropy of Hsp90 FRET pair 452-452 with AMPPNP. a) FRET efficiency vs. stoichiometry analysis reveals three conformational states of Hsp90: the open state ($E\sim$0.13), closed state A ($E\sim$0.63) and closed state B ($E\sim$0.85). To exclude donor-only and acceptor-only molecules exclusively single-molecule events with 0.3$<S<$0.7 were selected for further analysis. b) Schematic view of an anisotropy experiment. Fluorescence depolarisation depends on lifetime and rotation of the excited dye dipole and is measured by polarisation-sensitive detection. c) Acceptor anisotropy decay upon direct excitation at Hsp90 position 452 for closed state A. From a \textit{cone-in-cone} fit model we obtain rotational correlation times of 0.39$\pm$0.07\,ns for dye self-rotation, 2.7$\pm$0.7\,ns for local structural elements and 77$\pm$14\,ns for the global rotation. Plots and fit results for all subpopulations are shown in Fig.\,S8 and Tab.\,S4.}
	\label{fig:Aniso_452-452}
\end{figure}

To further clarify the origin of the observed time scale at $\sim$100\,ns we performed single-molecule time-resolved anisotropy experiments. Fig.\,\ref{fig:Aniso_452-452}a shows single-molecule fluorescence data of the FRET pair 452-452 with AMPPNP. Three conformational states can be distinguished by FRET efficiency $E$ vs. stoichiometry $S$ analysis : An open state at $E\sim$0.13, a closed state at $E\sim$0.63 and a more contracted closed state at $E\sim$0.85. This is consistent with previous results \autocite{Wolf.2021} and with the proximity ratios obtained in Fig.\,2. In the following the two closed states are referred to as closed states `A' and `B', respectively \autocite{Wolf.2021}. FRET efficiency and stoichiometry were used as selectors to perform subpopulation-specific anisotropy analysis. Fig.\,\ref{fig:Aniso_452-452}b shows the anisotropy decays for closed state A at position 452, respectively. To describe the data we used the \textit{cone-in-cone} model which accounts for free dye rotation $\rho_{\textrm{dye}}$, dynamics of the local environment $\rho_{\textrm{local}}$ and for global rotation of the overall protein $\rho_{\textrm{global}}$ (see Methods, Eq.\,\ref{eq:anisomodels})\autocite{Schroder.2005}.
A summary of all fit results is shown in Tab.\,S4.
We obtain $\rho_{\textrm{dye}}$ on the sub-ns time scale (0.39$\pm$0.07\,ns) which is comparable to the rotational correlation coefficient obtained for freely diffusing dyes \autocite{Vandenberk.2018}. For $\rho_{\textrm{local}}$, we obtain a correlation time of 2.7$\pm$0.7\,ns. The global rotation time $\rho_{\textrm{global}}$ occurs at 77$\pm$14\,ns. Most importantly, based on the weights of the dye and local component we can estimate the weight of global rotation to be below 1\,\%. Recalling that the 100\,ns component observed by nsFCS contributes with $\sim$10\,\%, we conclude that the 100\,ns time scale must be largely caused by internal dynamics of Hsp90 and not by rotation.

\clearpage
\subsection*{MD simulations show internal Hsp90 dynamics}

\textbf{MD simulations result in rotational correlation times similar to anisotropy decays}

\begin{figure}[h]
	\includegraphics[width=0.95\textwidth]{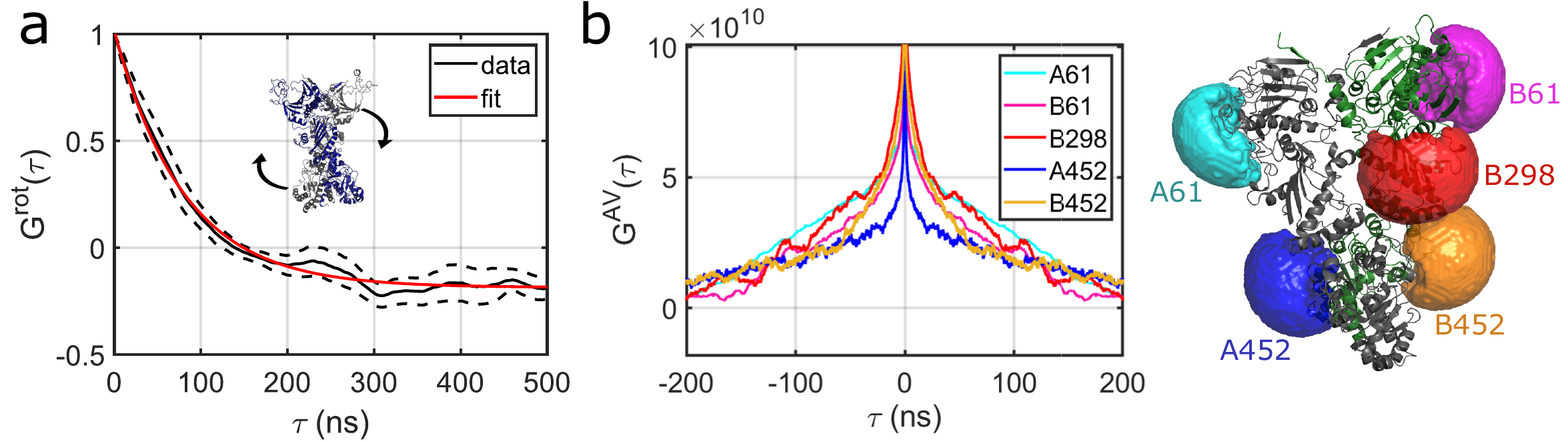}
	\caption{MD simulations provide a molecular mechanism for the nsFCS and anisotropy decay results. a) Autocorrelation of Hsp90's main axis of inertia orientation reveal a rotational correlation time of 81$\pm$1\,ns. b) Autocorrelation functions of accessible dye volumes at Hsp90 positions 61, 298 and 452 at chain A or B, respectively. Unconstrained bi-exponential fits reveal correlation times of around 100\,ns which hint to dye quenching by structural dynamics (see Fig.\,S9). A representative MD snapshot of the Hsp90 dimer visualises the investigated accessible dye volumes (colored spheres). Hsp90 chains A and B are coloured grey and green, respectively.}
	\label{fig:AVcorr}
\end{figure}

To establish a link between time-resolved anisotropy results and simulations, we characterized the rotational diffusion by calculating the autocorrelation function of the first principal axis of inertia. The movement of the axis of inertia is obtained from 5$\times$1\,$\upmu$s MD simulations of Hsp90 with AMPPNP as depicted in Figure \ref{fig:AVcorr}a. A single exponential fit reproduced the autocorrelation function poorly ($R^2 = 0.76$). We therefore use the equation
\begin{align}
   G^{\textrm{rot}}(\tau) = A \exp (\tau / \tau_1) + B \cos (2 \pi \omega \tau + \omega_0)
\end{align}
to account for contributions from potentially continuous rotation owing to simulation starting conditions that still persist after 2~$\upmu$s simulation time. The respective fit exhibits good agreement with $G^{\textrm{rot}}(\tau)$ ($R^2 = 0.99$) and results in a decay time constant \mbox{$\tau_1 = 81 \pm 1$~ns}, which ends up in excellent agreement with the experiment. Concerning the amount of persistent rotation, we find that the weight factors are $A = 1.17 \pm 0.02$ and $B = 0.199 \pm 0.002$. The ratio of rotational diffusion vs. continuous rotation is therefore about 6:1.

\textbf{MD-based accessible dye volume correlations reveal structural Hsp90 dynamics}

Complementary to time-resolved anisotropy experiments, which selectively probe rotational dynamics, MD simulations can provide access to local dynamics of Hsp90 which were disentangled from global rotational diffusion. Therefore, we monitored the change of the accessible dye volumes (AV) over time at the same Hsp90 positions which were investigated by nsFCS, namely amino acid numbers 61, 298 and 452 (see Fig.\,\ref{fig:AVcorr}b for an illustration). To investigate position-specific AV changes on characteristic time scales, we calculated the autocorrelation functions of the AV (Fig.\ref{fig:AVcorr}b).
We obtain correlation curves exhibiting two dominant decays. Unconstrained bi-exponential fits reveal a fast correlation below 100\,ps, which is faster than the minimum lag time and therefore not further discussed here. More interestingly, all positions show a slower correlation on the 100\,ns time scale (see Fig.\ref{fig:AVcorr}b, Fig.\,S9 and Tab.\,S6 for summary of all fits).
A likely scenario, which explains this observation, is that structural fluctuations confine the dye flexibility. From the fluorescence viewpoint a smaller accessible dye volume translates into a higher probability of collision with neighboring side chains and thus a higher possibility of quenching. These observations provide a direct link between nsFCS and theory and support our hypotheses that structural dynamics contribute to the observed $\sim$100\,ns nsFCS time scale.
Interestingly AV cross-correlations show an anticorrelated component with amplitudes which are about one order of magnitude weaker than those of the AV autocorrelations and about the same time scale of 100\,ns (Fig.\,S9b). In the experiment we detect a superposition of these signals, which likely explains why we do not see the anti-correlated signal there. From this, we conclude that combining nsFCS with information from time-resolved anisotropy and MD simulations is a versatile strategy to disentangle rotational from internal dynamics which puts us in the ideal situation to compare them to results from complementary techniques such as neutron spin echo.

\textbf{Cartesian PCA reveals molecule-spanning dynamics of the full Hsp90 dimer}

We now address a cartesian principal component analysis of all backbone and C$_{\beta}$ atoms from one representative of the five 1~$\upmu$s simulations (Fig.\,S10). The eight eigenvectors that cover 80\,\% of the overall positional variance all represent molecule-spanning motions of the full dimer, as can be seen from the root mean square fluctuation (RMSF) per atom and eigenvector along the protein chain. The time traces of projections per eigenvector and its autocorrelation function (ACF) show that no clear separation between slow or fast modes exist. Instead, a continuum of time scales between tens and several hundreds of nanoseconds appears.
This continuum does not stand in contrast to the well-defined time scale observed in nsFCS. The fluorescence correlation time depends on the time scale of changes in dye-accessible volumes, which represents more localized structural changes, while we extract global structural changes here. The first two eigenvectors (1 and 2) contain the majority of variance in motions ($\approx$60\,\% of total variance), i.e., are the largest motions. The ACF analysis reveals that they constitute the slowest motions, as well.
The respective structures of these projections along eigenvectors 1 and 2 over time are shown in Supplementary Movies 1 to 4. While morphing between the states with maximal projection values (Supplementary Movies 1 and 2) highlights the delocalized nature of these motions, we want to emphasize that the eigenvectors calculated here do not encode low frequency vibrations, but a highly diffusive and fluctuating bending-contraction motion, as can be seen in the actual time series of projections (Supplementary Movies 3 and 4).

\newpage

\textbf{Diffusion from rigid body calculations}

In order to validate the simulation results in terms of absolute time scales, we estimated the translational and rotational diffusion coefficient for the average protein structures of the five independent MD runs. To this end, we employed hydrodynamic bead modeling via the software package HYDROPRO\autocite{Ortega2011} resulting in the diffusion tensor of a rigid body, i.e.~neglecting any internal dynamics. The resulting translational diffusion coefficient of $D_t^\text{(rigid)} = (4.71\pm 0.04) \cdot 10^{-7}$~cm$^2$/s (20 $^\circ$C in water) is in good agreement with the translational diffusion coefficient $D_t^\text{(MD)} =(4.4 \pm 0.5) \cdot 10^{-7}$~cm$^2$/s from the MD simulations. The rotational diffusion coefficient yields $D_r^\text{(rigid)} = (1.64\pm 0.03) \cdot 10^{6}$/s (20 $^\circ$C in water). While a direct comparison to the rotational diffusion of the main axis is not straight-forward, the theoretical sequence of rotational relaxation times $\tau_1=1/2D_r\approx300$\,ns and $\tau_2=1/6D_r\approx100$\,ns have the right order of magnitude. This good agreement supports both the validity of the MD simulations, and stresses that rotational contributions are expected on time scales around 100 ns.
\clearpage

\subsection*{NSE shows a global internal diffusive mode on nanosecond time scales}

\begin{figure}
    \includegraphics[width=\textwidth]{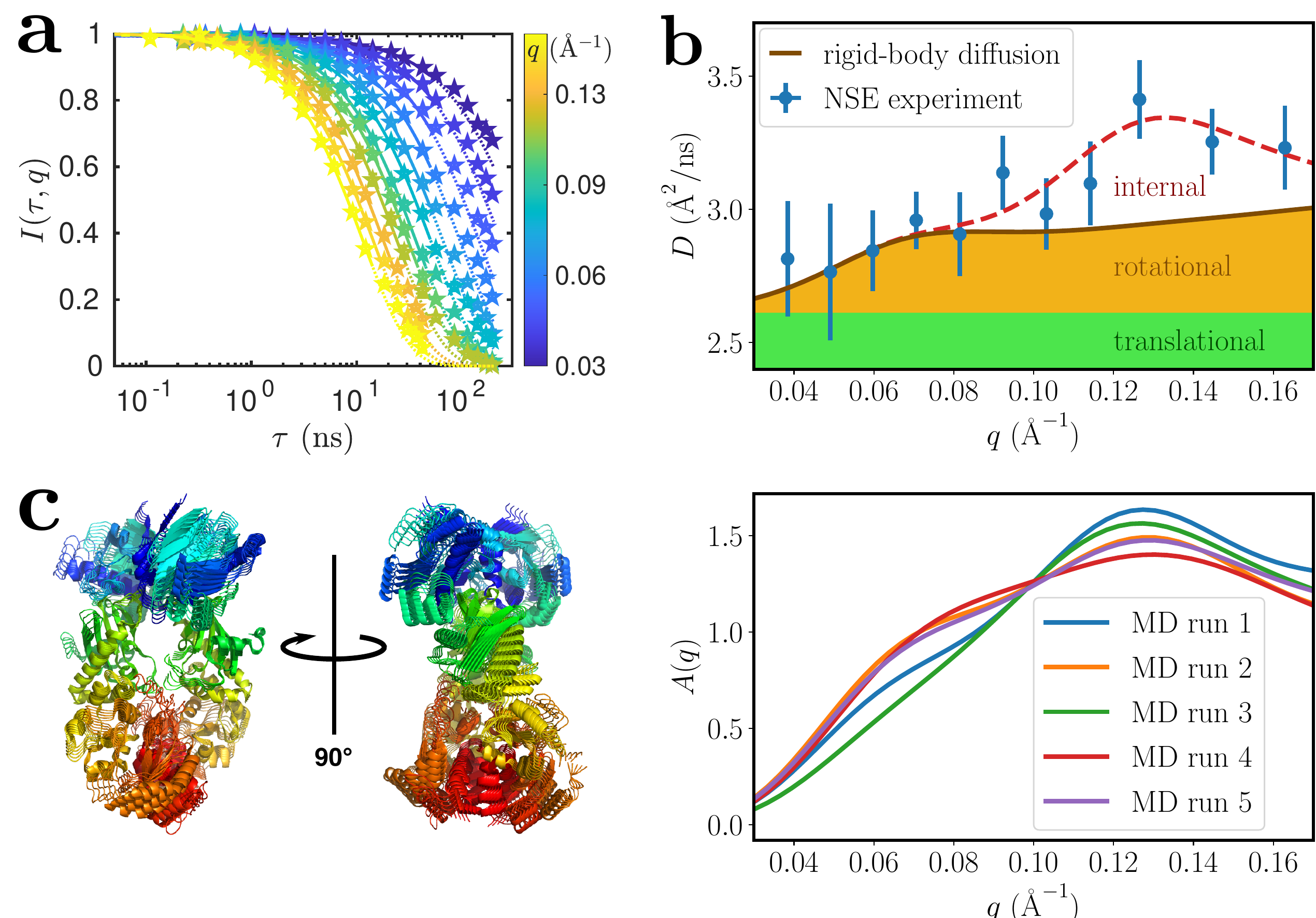}
\caption{Neutron spin echo spectroscopy shows internal motions beyond the translation and rotation of the entire molecule. a) Fits of the intermediate scattering functions for different scattering vectors $q$ (color coded). The fits (solid lines) were performed using Equation \ref{eq:NSE:Fit} for $\tau<30~\textrm{ns}$ and for $I(q,\tau)>0.3$. They were extrapolated with dotted lines to outline the expected deviation from the single exponential suggesting the presence of internal motions.  b) The resulting experimental diffusion function $D(q)$ evidences a first shoulder around the scattering vector $q\approx 0.07$\,\AA$^{-1}$ and a peak around $q\approx 0.13$\,\AA$^{-1}$. While the shoulder can be described by rotational diffusion of the entire protein based on a rigid-body modeling, the peak originates from an internal degree of freedom. c) Visualization of the first and most relevant principal component of Hsp90: a bending motion with a slight twist spans over the entire molecule (left). Calculated signatures of the ten PCA eigenvectors obtained from five MD runs with independent start parameters. The peak around $q\approx 0.13$\,\AA$^{-1}$ indicates that Hsp90 performs concerted internal motions on times scales of 10-100 ns (right), consistent with nsFCS.}
\label{fig:NSE}
\end{figure}

NSE experiments provide unique information on the collective motions inside proteins \autocite{Biehl2011, Bu.2011, Grimaldo_2019_QuartRevBiophys}. The initial decay of the experimental intermediate scattering function $I(q,\tau)$ (Fig.~\ref{fig:NSE}a) was determined by a single exponential fit to the initial slope ($\tau<30$\,ns and $I(q,\tau)>0.3$) for each value of the scattering vector $q$ individually:
\begin{equation}
    I(q,\tau)=C\cdot\exp\left(-D_\textrm{eff}q^2\tau\right)
    \label{eq:NSE:Fit}
\end{equation}
with a scalar prefactor $C\approx 1$.
The resulting effective short-time diffusion function $D_\textrm{eff}(q)$ (Fig.\,\ref{fig:NSE}b) contains contributions from the translational and rotational diffusion of the entire protein, as well as signatures of internal diffusive modes such as multi-domain dynamics and bending \autocite{Biehl_2014,Grimaldo_2019_QuartRevBiophys}.
Two main features are observed. At lower $q\approx 0.07$\,\AA$^{-1}$ corresponding to length scales of $\sim$10\,nm, $D_\textrm{eff}(q)$ shows a shoulder. At higher $q\approx 0.13$\,\AA$^{-1}$ corresponding to length scales of $\sim$4.8\,nm, a peak is observed.

In order to model the different contributions, we started from the effects arising from rigid-body translational and rotational diffusion (solid line in Fig\,\ref{fig:NSE}b, for details see SI). The absolute values of $D_\textrm{eff}$ start at the translational diffusion coefficient in the low $q$ limit, and then increase due to contributions of rotational diffusion on $q$ values corresponding to the entire protein size. We remark that the absolute value of $D_\textrm{eff}$ is incompatible with a pure dimer solution (see Fig.\,S11). We used a hypothetical hexamer solution by rescaling the translational and rotational diffusion coefficient based on their relation to the protein radius as $D_t^{(hex)} = D_t^{(dimer)}/\sqrt[3]{3}$ and $D_r^{(hex)} = D_r^{(dimer)}/{3}$. Note that the presence of some amount of oligomers is also evident from the pair distribution functions $p(r)$ for the identical and for the similar samples calculated from SANS measurements (see Fig.\,S12).

Indeed, this approximative modeling recovers the shoulder well, but clearly fails to explain the peak feature at $q\approx 0.13$\,\AA$^{-1}$, and in general higher values of $D_\textrm{eff}(q)$ at larger $q$. This deviation is shown as dashed guide to the eye in Fig.\,\ref{fig:NSE}b and is indicative of additional internal motions of the protein, as they occur at $q$ values corresponding to motions within the protein. We stress that the significance of this additional contribution is independent of the assumption of oligomers.

For more detailed modeling of the internal motions, we based calculations on the cartesian PC analysis from the MD simulations. As the central observable, an amplitude function $A_k(q)$ is obtained for the $k$th eigenvector:\autocite{Biehl2011,Grimaldo_2019_QuartRevBiophys}
\begin{equation}
A_k(q) = \sum\limits_{\alpha,\beta} b_\alpha b_\beta \exp\left(i\mathbf{q}\cdot(\mathbf{r}_\alpha-\mathbf{r}_\beta)\right)(\hat{\mathbf{q}}\cdot\mathbf{\hat v}_{k\alpha})(\hat{\mathbf{q}}\cdot\mathbf{\hat v}_{k\beta})
\end{equation}
where the sum runs over all atom pairs with index $\alpha$ and $\beta$ with coordinates $\mathbf{r_x}$. $b_x$ denote scattering cross-sections of the atoms. The cartesian PC eigenvector provides displacements vectors $\mathbf{\hat v_{kx}}$ at each atom position, which together with the scattering vector $\mathbf{q}$ allow for the geometric interpretation of collective structural dynamics of multi-domain proteins. The first ten PC eigenvectors were used for further analysis.
We weighted the amplitudes according to the square-root of the corresponding PC eigenvalues, and finally obtained the expected experimental signature of the internal motions present in MD simulations (Fig.\,\ref{fig:NSE}c). Importantly, all five independent runs of MD simulations show a consistent peak at $q\approx 0.13$\,\AA$^{-1}$, which unambiguously indicate internal motions with correlation lengths of around 4.5 nm, i.e.~spanning over a large part of the molecule.

We remark that our analysis considers all dynamical contributions in one cumulant exponential term, so that no specific value of the internal relaxation time can be assigned to the slowest mode. However, the decays of the correlation functions (Fig.\,\ref{fig:NSE}a) occur on time scales between 10-100 ns, which provide a good estimate for the related relaxation time consistent with the previous nsFCS and MD simulation results.

As an independent validation of the absolute diffusion coefficients and thus the presented NSE analysis, we used neutron backscattering spectroscopy (NBS) which is expected to provide comparable diffusion coefficients in the high $q$ limit of NSE.
Following the conventional analysis (Fig.\,S13b) \autocite{Grimaldo_2019_QuartRevBiophys}, we obtain information on both diffusion of the entire protein, and the geometrical confinement of small local motions down to sub-nanosecond time scales.
We obtain an apparent global diffusion coefficient of $D_\textrm{app}=(3.27 \pm 0.18)$\,\AA$^{2}$/ns, consisting of the translational and rotational short-time self diffusion coefficient\autocite{Roosen-Runge2011}, which is in excellent agreement with the NSE data.
In addition, the confinement radius of local motions results in $a = (1.95 \pm 0.07)$\,\AA, which is fully consistent with the length of side chains limiting their motions. In particular, this finding stresses that local motions cannot explain the peak at $q\approx 0.13$\,\AA$^{-1}$, and larger concerted motions need to be at play.

\clearpage
\subsection*{The co-chaperone Sba1 affects nanosecond dynamics of Hsp90}

\textbf{Single-molecule fluorescence shows decelerated Hsp90 dynamics in presence of Sba1}

We performed single-molecule fluorescence experiments of Hsp90 with AMPPNP in presence and absence of Sba1 to investigate its effect on the nanosecond dynamics.

As Hsp90 label site we chose position 298 which is slightly offset with respect to the Sba1-binding site in the crystal structure 2cg9 \autocite{Ali.2006}. Here we chose position 298, as it is closer to the Sba1-binding site compared to position 452. Atto532 was used as fluorescent label, Sba1 was added without label.

\begin{figure}[ht]
	\includegraphics[width=0.9\textwidth]{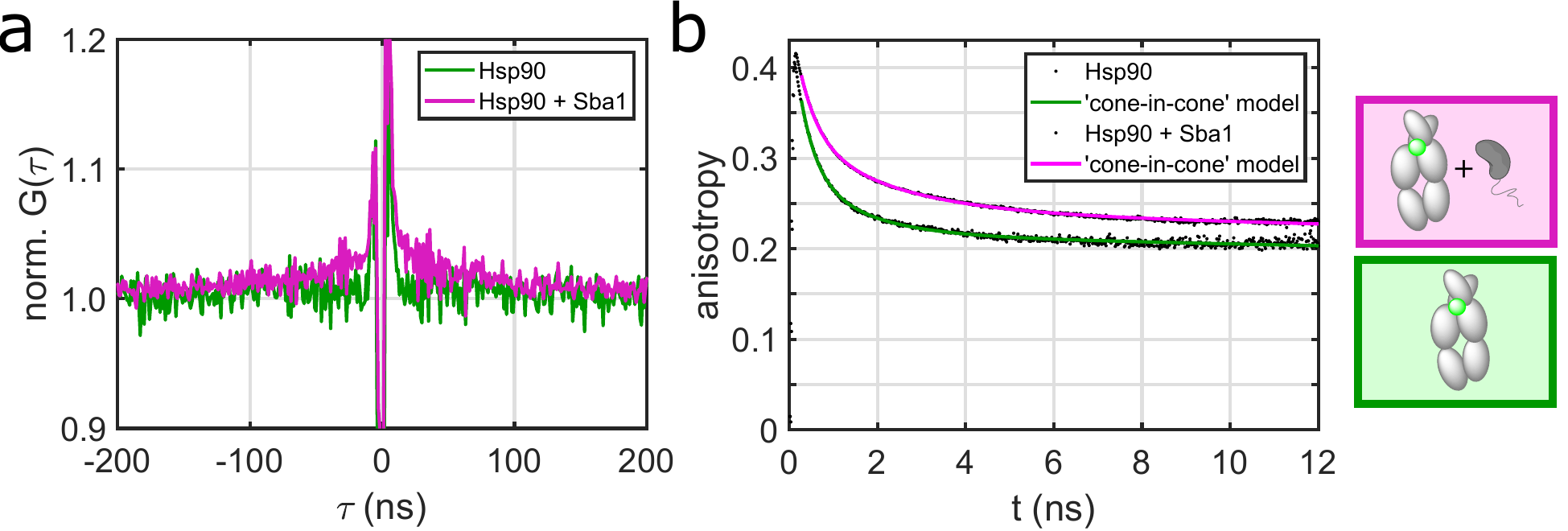}
	\caption{The co-chaperone Sba1 modulates Hsp90 dynamics on the nanosecond time scale. Hsp90 was singly labeled with Atto532 at position 298 and measured with AMPPNP. Sba1 was added unlabeled. a) nsFCS data in absence (green) and presence of Sba1 (magenta). With Sba1 the bunching time of Hsp90 decelerates from 65$\pm$22\,ns to 100$\pm$18\,ns. See Tab.\,S2 for all fit results. b)~Anisotropy decays and \textit{cone-in-cone} fit models of Hsp90 in absence (green) and presence of Sba1 (magenta). Upon addition of Sba1, the local rotation slows down from 1.4$\pm$0.3\,ns to 2.2$\pm$0.2\,ns. The increased residual anisotropy indicates binding of Sba1 to Hsp90. For details on the fit see Tab.\,S5.}
	\label{fig:Hsp90-Sba1}
\end{figure}

The nsFCS results are shown in Fig.\,\ref{fig:Hsp90-Sba1}a. The data is well described using one antibunching and two bunching components (Eq.\,\ref{eq:nsFCSmodel}b, SI). Details on the fit results are shown in Tab.\,S2. The fast bunching mode is obtained at 5.2$\pm$0.2\,ns and can therefore consistently be attributed to tryptophan quenching. Given the label position, this component is expectedly unaffected by the presence of Sba1. In contrast, with Sba1 the time scale of the slower bunching mode shifts from 65$\pm$22\,ns to 100$\pm$18\,ns.

Supporting information was obtained by anisotropy decay analysis (see Fig.\,\ref{fig:Hsp90-Sba1}b). In the presence of Sba1 the residual anisotropy is clearly increased, which indicates the presence of Hsp90-Sba1 complexes. To obtain information on local dynamics, in line with our previous analysis we applied the cone-in-cone fit model \autocite{Schroder.2005} (See Eq.\,\ref{eq:anisomodels} (SI) and Tab.\,S5 for fit results). Here, due to weaker statistics we constrained the maximum global rotational correlation time $\rho_{\textrm{global}}$ to 200\,ns which implies the reasonable assumption that the protein complexes formed did not exceed molecular weights of 500\,kDa \autocite{Lakowicz.2013}.
This way, upon additon of Sba1, we observe a shift of the local rotational correlation time of Hsp90 ($\rho_{\textrm{local}}$) from 1.4$\pm$0.3\,ns to 2.2$\pm$0.2\,ns. Dye self-rotation was obtained at 0.40$\pm$0.06\,ns and 0.45$\pm$0.03\,ns in the absence and presence of Sba1, respectively. Note that the change in local rotational dynamics cannot be an artefact from changed instrumental response functions because anisotropy decays with and without Sba1 were recorded at the same detectors with comparable count rates. Furthermore, the fact that $\rho_{\textrm{dye}}$ is consistent with reported values for free dye rotation\autocite{Vandenberk.2018} shows that the effect of the instrument response is weak and justifies to circumvent the often error-proned reconvolution analysis. Both, nsFCS and anisotropy decay experiments independently show that Sba1 decelerates the local nanosecond dynamics of Hsp90.

\textbf{Anisotropic network models imply a reduction of Hsp90 dynamics by Sba1 around the N/M interface}

The results from anisotropic network modeling (ANM) analysis for both Sba1-bound and unbound Hsp90 are displayed in Fig.~S15. In ANM, small eigenvalues approximately indicate slow normal mode frequencies. Similar to the results from the cartesian PCA, slow modes represent global motions of the full dimer. The eigenvalue distribution itself does not exhibit any difference between the two states, so the frequency distribution of Hsp90 appears to remain unchanged upon Sba1 binding in this simple model. However, individual normal modes exhibit locally changed dynamics: eigenmode 4 is the slowest mode predicted with a change in dynamics around the binding site of Sba1. For comparison, we show eigenmode 2, which represents a similar twist motion, but without effect of Sba1. Sba1 therefore likely locally dampens motions of the dimer, which is in agreement with our results from nsFCS and time-resolved anisotropy experiments.

\clearpage
\section*{Conclusion}

Combining advanced fluorescence techniques, quasi-elastic neutron scattering and full-atom MD simulations, we shed light on the previously unaddressed nanosecond dynamics of the molecular chaperone Hsp90 (see Fig.\,\ref{fig:timescale-overview}).
The complementary use of these methods rooted in bio- and soft matter physics offers unique insights because temporal and spatial information is obtained simultaneously in a coupled manner. Furthermore, the specificity of the techniques employed allowed us to distinguish internal dynamics from global rotation, although occurring at very similar time scales.

\begin{figure}[h]
	\centering
	\includegraphics[width=0.7\textwidth]{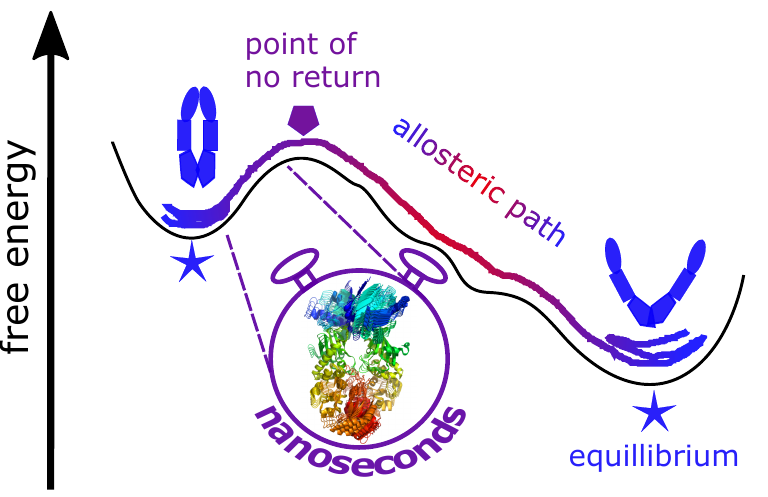}
	\caption{Molecule-spanning dynamics on the 100\,ns time scale are precursors of large conformational changes between two equilibrium states (marked with a star). They probe higher regions of the free energy landscape and likely guide the protein towards the point of no return (indicated by the pentagon) to complete e.g. an allosteric signalling path.}
	\label{fig:funcrel}
\end{figure}

Most interestingly, we observe molecule-spanning dynamics that are distributed across the complete protein on the 100\,ns time scale. This internal dynamics can be described as a diffusive motion, in the present case a bending contraction along the N-M axis of the closed Hsp90 dimer. This motion cannot be described by a simple harmonic oscillator anymore. Therefore, we are convinced that we do not only probe the minimum of the free energy of this state, but already explore higher regions of the free energy landscape, which likely is a precursor for larger conformational changes within this multi-domain protein (see Fig.\,\ref{fig:funcrel}). 

We further showed that these nanosecond dynamics of Hsp90 can be affected by a co-chaperone, Sba1. We observe that Sba1 decelerates Hsp90 nanosecond dynamics which could facilitate its interaction with clients by populating the binding-competent state for longer times and therefore making it more accessible. Altogether, our study indicates that molecule-spanning nanosecond dynamics are the precursor for large conformational changes and that they constitute a previously underestimated and biologically important level of protein regulation.

\clearpage
\section*{Material and Methods}
\subsection*{Biochemistry and sample preparation}

\textbf{Protein production and purification.}
Gene expression and subsequent protein purification was performed as previously described \autocite{Schmid.2020}.
Yeast Hsp90 wildtype and variants with cysteine mutations at positions 61, 298 and 452 for fluorescent labeling were recombinantly produced in \textit{E. coli}.
Protein purification was performed with three consecutive chromatography steps, Ni-affinity, anion-exchange and size exclusion chromatography, yielding nearly monodisperse dimeric Hsp90. Altogether about 1\,g of protein was purified for all the experiments. A detailed description is given in the SI.\\

\textbf{Fluorescent Labeling and single-molecule experiments.}
Fluorescent labeling was achieved by site-directed cysteine-maleimide chemistry. Atto532 or Atto550 were used as donor and Atto647N as acceptor, respectively (ATTO-TEC GmbH). Note donor- and acceptor positions of a FRET pair are always specified in the order \lq donor-acceptor\rq.
To stabilise the Hsp90 dimer at single-molecule concentration a coiled-coil motif (DmKHC, \textit{D. melanogaster}) was inserted at the C-terminus which prevents dimer dissociation. Hsp90 heterodimers with one donor and one acceptor molecule per dimer were obtained by incubating donor- and acceptor-labeled Hsp90 dimers for 45\,min at 43$^{\circ}$C at ratio 1:1. At this temperature monomers exchanged which yielded stochastically one half Hsp90 heterodimers. The samples were centrifuged for 1\,h at 4$^{\circ}$C to separate potential aggregates.
To obtain Hsp90 dimers with only one dye, donor-labeled Hsp90 dimers were exchanged with Hsp90 wildtype dimers at a 1:10 ratio.
As buffer 40\,mM HEPES, 150\,mM KCl and 10\,mM MgCl$_2$ were dissolved in ultra pure H$_2$O and pH was adjusted to 7.5. Final protein concentrations were $\sim$100\,pM for single-molecule and $\sim$200-500\,pM for nsFCS experiments. AMPPNP was added to the samples immediately before the start of the measurement such that the final nucleotide concentration was 2\,mM.\\

\textbf{Sample preparation for neutron scattering experiments.}
In all neutron experiments the yeast Hsp90 wildtype construct without the coiled-coil motif and without cysteine mutation was measured. A buffer exchange from H$_2$O-based buffer to 150\,mM KCl and 10\,mM MgCl$_2$ dissolved in D$_2$O was achieved by five consecutive dialysis steps at 8$^{\circ}$C over night (Slide-A-Lyzer 10K, Thermo Scientific). To ensure that the signal of free H$_2$O is reduced sufficiently, a D$_2$O-based buffer excess of $>$50:1 was applied in each dialysis step. Directly before the measurement samples were centrifuged 20\,min at 12857\,g and 8$^{\circ}$C. Final protein concentrations determined by UVvis spectroscopy (Nanodrop ND-1000, Thermo Fisher Scientific) were 614\,$\upmu$M. \\

\textbf{Check of bio-functionality}
Bio-functionality of the Hsp90 dimer was controlled according to well-established ATPase assays \autocite{Richter.2001}. The assay couples ATP hydrolyzation to the decrease of NADH which was monitored over time on a Lambda35 UVvis spectrometer. After addition of 2\,mM ATP and 2\,$\upmu$M Hsp90 a linear decrease of the absorbance at 340\,nm was obtained (see Fig.\,S3). In addition we validated the open-close dynamics of Hsp90 in D$_2$O-based buffer by single-molecule experiments (see Fig.\,S4).

\subsection*{Fluorescence experiments}
Single-molecule fluorescence experiments were conducted on a home-build confocal setup as depicted in Fig.\,S1. Green (532\,nm, LDH-D-FA-530L) and/or a red laser light (640\,nm, LDH-D-C-640, PicoQuant) were used to excite donor and acceptor molecules. Before focusing on the sample by a 60x water immersion objective (CFI Plan Apo VC 60XC/1.2 WI, Nikon), both beams were polarized and overlaid by a dichroic mirror (zt 532 RDC, AHF). In the emission path a second dichroic mirror (F53-534 Dual Line beam splitter z 532/633, AHF) separated donor fluorescence from acceptor fluorescence. Pinholes (150\,$\upmu$m diameter) filtered off-focus light. Before detection, polarizing beam splitters separated parallel and perpendicular polarized light. Green and red emission was detected by single-photon detectors (two SPCM-AQR-14, PerkinElmer and two PDM series APDs, Micro Photon Devices).

\textbf{Time-resolved anisotropy.}
To obtain time-resolved single-molecule anisotropy decays pulsed-interleaved excitation was used to alternately excite donor and acceptor molecules at 20\,MHz repetition rate. Excitation powers directly before the objective were 317\,$\upmu$W and 114\,$\upmu$W for green and red, respectively. Microtimes and macrotimes were recorded in T3 mode by time-correlated single photon counting (HydraHarp400, PicoQuant) with 50\,ns and 16\,ps time-resolution, respectively.
Data were analysed in Matlab by the software PAM and self-written scripts.
Single-molecule events where identified using a search algorithm with threshold criterion of at least 100\,photons per event. Photon traces were chopped into 1\,ms bins and photons were ascribed to a single-molecule event if the bin count rate exceeded 80\,kHz.
For each single-molecule event FRET efficiency $E$, stoichiometry $S$ and anisotropy $r$ were calculated according to Eq.\,\ref{eq:E}, Eq.\,\ref{eq:S} and Eq.\,\ref{eq:aniso}, respectively.
\begin{equation}
\label{eq:E}
E = \dfrac{GF_{DA} -\alpha F_{DD}-\delta F_{AA}}{\gamma F_{DD} + F_{DA}-\alpha F_{DD}-\delta F_{AA}}
\end{equation}
Here, $F_{ij}$ stands for the detected fluorescence intensities which are detected in channel $i$ after excitation of $j$ ($i, j = \{$D, A$\}$, with D = donor, A = acceptor), respectively. $E$ is corrected for crosstalk by $\alpha$, for direct excitation by $\delta$ and for local differences in quantum yield and detection efficiency by $\gamma$. The G-factor corrects for differences between parallel and perpendicular detection channels.

\begin{equation}\label{eq:S}
S = \dfrac{\gamma F_{DD} + F_{DA}-\alpha F_{DD}-\delta F_{AA}}{\gamma F_{DD} + F_{DA}-\alpha F_{DD}-\delta F_{AA} + \beta F_{AA}}
\end{equation}

In addition to the correction factors $\alpha$, $\delta$ and $\gamma$, the stoichiometry is corrected by $\beta$ to account for differences in absorption cross-section and excitation powers of donor and acceptor fluorophores, respectively.

\begin{equation}\label{eq:aniso}
r = \dfrac{GF_{\parallel}-F_{\perp}}{(1-3l_2)GF_{\parallel}+(2-3l_1)F_{\perp}}
\end{equation}

Here, subscripts of the fluorescence intensities $F$ denote the polarisation channel. $l_1$ and $l_2$ account for the depolarisation effect of the objective. We used $E$ and $S$ to filter and histogram microtimes belonging to a specific conformational state.
The subpopulation-specific anisotropy decays were analysed using the \textit{cone-in-cone} model \autocite{Schroder.2005}:\\
\begin{equation}\label{eq:anisomodels}
r(t) = r_0((1-A_{\mathrm{dye}})e^{-t/\rho_{\textrm{dye}}}+A_{\textrm{dye}})((1-A_{\textrm{local}})e^{-t/\rho_{\textrm{local}}}+A_{\textrm{local}})e^{-t/\rho_{\textrm{global}}}
\end{equation}
Here, $\rho_{\textrm{dye}}$ accounts for correlations due to dye self-rotation, $\rho_{\textrm{local}}$ for rotational dynamics of local structural elements the dye is attached to and $\rho_{\textrm{global}}$ for rotation of the overall protein. Note that a reconvolution analysis was not necessary because the time scale of interest ($>$1\,ns) was well separated from the time scale of instrument response.

\textbf{Subpopulation-specific nanosecond FCS.}
Green or red cw laser excitation was used to generate nsFCS data. The laser powers were 84\,$\upmu$W, 287\,$\upmu$W or 138\,$\upmu$W for Atto532-, Atto550- and Atto647N-labeled Hsp90, respectively. Additional IR filter (LC-HSP750-25, LaserComponents) were inserted before the detectors to reduce artefacts from detector afterglowing\autocite{Kurtsiefer.2001}. Photon arrival times were saved in T2 mode (HydraHarp400, PicoQuant) which provided a time-resolution of 1\,ps. nsFCS data analysis was performed using the Mathematica package Fretica (Ben Schuler group). Single-molecule events were identified using the $\Delta$\,T-search algorithm with parameters $\Delta$\,T=50\,$\upmu$s, N$_{min}$=50 and N$_{max}$=1000. Subpopulation-specific analysis was achieved based on the proximity ratio $PR$ according to Eq.\,\ref{eq:PR}
\begin{equation}\label{eq:PR}
	PR = \dfrac{F_{DA}}{F_{DD}+F_{DA}}
\end{equation}
Subscripts of the fluorescence intensities $F$ denote the respective detection channels with the same syntax as above. To investigate only events in which Hsp90 is in a closed we analysed only data with $PR>0.3$. Note that the proximity ratio is the uncorrected form of the FRET efficiency $E$ and therefore contains all information required for subpopulation-specific analysis.

Auto- and cross-correlations were calculated for donor-donor (Don$\times$Don), acceptor-acceptor (FRET$\times$FRET) and donor-acceptor signals (FRET$\times$Don), respectively. For Don$\times$Don and FRET$\times$FRET parallel and perpendicular channel signals were cross-correlated to avoid artefacts resulting from afterpulsing. Correlations were calculated on a time window of 500\,ns with a lag time of 1\,ns. Shown nsFCS data of Hsp90 were described according to Eq.\,\ref{eq:nsFCSmodel}.
\begin{subequations}\label{eq:nsFCSmodel}
	\begin{alignat}{2}
	G(\tau) &= a(1-c_{ab}e^{-(\tau-\tau_0)/\tau_{ab}}) (1+c_{b1}e^{-(\tau-\tau_0)/\tau_{b1}}) \\
	G(\tau) &= a(1-c_{ab}e^{-(\tau-\tau_0)/\tau_{ab}}) (1+c_{b1}e^{-(\tau-\tau_0)/\tau_{b1}}) (1+c_{b2}e^{-(\tau-\tau_0)/\tau_{b2}})
	\end{alignat}
\end{subequations}
The fit models include a scaling factor $a$, the weight of the antibunching mode $c_{ab}$, the weight of one or two bunching modes $c_{b1}$ and $c_{b2}$, the antibunching time $\tau_{ab}$ and one or two bunching times $\tau_{b1}$ and $\tau_{b2}$, respectively. $\tau_0$ corrects for a small delay of the detection channels. Note that $\tau_{b1}$ is a global fit parameter which was used to describe Don$\times$Don, FRET$\times$FRET and FRET$\times$Don correlation data simultaneously.
Please refer to the SI Methods for details on all investigated FRET pairs, important cross-checks on the effect of Trp quenching and single-labeled species (Fig.\,S4-S6). A summary of all fit results is given in Tab.\,S1-S3.

\subsection*{Neutron scattering experiments}
Neutron scattering offers non-destructive access to the nano-scale. Different techniques provide access to the structural and dynamical properties of the investigated samples at different time scales reaching from dynamics on the pico-second time scale to static structures\autocite{Grimaldo_2019_QuartRevBiophys}.
The coherent and incoherent isotope-dependent scattering cross-sections lead to the coherent and incoherent scattering signal, which sum up to the total scattering signal.
The fraction of the different contributions depends on the investigated scattering angle $2\theta$ which together with the wavelength $\lambda$ can be linked to the momentum transfer~ $\hbar\textbf{q}$ whereby $q=\left|\mathbf{q}\right|$ is the absolute value of the scattering vector:
\begin{equation}
    q=\frac{4\pi \sin\left(\frac{2\theta}{2}\right)}{\lambda}.
\end{equation}

\textbf{Neutron Spin Echo Spectroscopy.}
NSE experiments were performed at the spin-echo spectrometer IN15 (ILL, Grenoble)\autocite{Farago2015}. Samples were measured at 50\,mg/ml in quartz cuvettes (2\,mm thickness) at 295K. Four different detector angles were configured with different wavelengths ($2\Theta=8.3^\circ,\lambda=6\textrm{\AA}$; $2\Theta=3.5^\circ,\lambda=10\textrm{\AA}$; $2\Theta=6.5^\circ,\lambda=10\textrm{\AA}$; $2\Theta=9.5^\circ,\lambda=10\textrm{\AA}$).
The position sensitive detector was subdivided in three $q$ values, resulting in twelve q-values between $0.028~\textrm{\AA}^{-1}<q<0.16~\textrm{\AA}^{-1}$.
Standard methods were applied for data reduction and background subtraction as detailed in Ref. \cite{Hoffmann2021,Mezei1980a}.

\textbf{Quasi-Elastic Neutron Backscattering.}
NBS data were measured at the backscattering spectrometer IN16B (ILL, Grenoble). The sample measured by NSE and NBS was identical which ensured the best data consistency. To this end, the 50\,mg/ml Hsp90 solution was transferred from the NSE quartz cuvette to a cylindrical aluminium holder with a 0.15\,mm gap between inner and outer radius.
Unpolished Si(111) Analyzers were used in combination with a chopper ratio 1:1 ('high flux' mode) and with the Doppler monochromator.  D$_2$O buffer and vanadium were measured for 4\,h to correct for background signal and to determine the energy resolution, respectively.

\subsection*{Computational methods}

MD simulations are prolongation runs of the 5x 1$\upmu$s simulation runs a AMPPNP-bound Hsp90 dimer in the closed conformation presented in Ref.~\cite{Wolf.2021}. In brief, simulation systems were generated from PDB ID 2CG9 \autocite{Ali.2006} and simulated using Gromacs v2020 \autocite{Abraham.2015} with the Amber99SB*ILDN-parmbsc0-$\chi\, OL3$ + AMBER99ATP/ADP force field. \autocite{PerezVilla.2015} We continued the respective simulations to a further 5x 1$\upmu$s and treated the initial 1$\upmu$s from our earlier work as equilibration period. Coordinates were saved each 100~ps.

Accessible dye volumes were calculated based on structural MD snapshots using the Olga software \autocite{Kalinin.2012,Dimura.2019}. AVs were calculated based on the equilibrated 5$\times$1\,$\upmu$s MD traces of the Hsp90-AMPPNP simulations. This resulted in 10000 accessible volumes with a time resolution of 100\,ps averaged over five independent MD runs.
Translational diffusion coefficients and principal axes of inertia were calculated with Gromacs-internal tools. The rotation autocorrelation function $G^{\textrm{rot}}(\tau)$ was calculated from the first principal axis of inertia's eigenvector $\mathbf{I}$ cartesian $i = x,y,z$ components $I_i$
\begin{align}
   G^{\textrm{rot}}(\tau) = \frac{1}{3N} \sum_{i = x,y,z}^i \sum_j^N \frac{ \left( I_i (t_j + \tau) - \left< I_i \right> \right)  \left( I_i (t_j) - \left< I_i \right> \right)}{\sigma_{I_i}^2}
\end{align}
for all $N$ discrete time steps $t_j$ that can be overlapped with the original time series within a shift $\tau$, time series means $\left< I_i (t_j) \right> $ and variance $\sigma_{I_i}^2$. To remove artifacts in the calculation $G^{\textrm{rot}}(\tau)$ owing to random sign flips in the $\mathbf{I}$ time series, we flipped the sign of $\mathbf{I}$ in case that any $I_i$ changed by more than 0.5 units between two time steps.

The cartesian principal component analysis (cPCA)\autocite{Sittel.2014} was carried out using Gromacs-internal tools independently for all five AMPPNP trajectories. We excluded the charged loop between residue numbers 208 and 280 from analysis due to its high flexibility and according dominance of the first principal components. After rotational and translational fit of the protein C$_\alpha$ atoms to the final protein structure, the mass-weighed cartesian covariance matrix of the protein backbone and adjacent C$_\beta$ atoms was calculated and diagonalized. The resulting eigenvectors and eigenvalues were subsequently used for analysis in the analysis of neutron scattering experiments.

The anisotropic network model (ANM) analysis was performed using the ANM web server v2.1. \autocite{Eyal.2015} As in simulations, we used PDB ID 2CG9 as input. For analysis of the effect of the Sba1 co-chaperone on the Hsp90 dimer, we created one structure with only one Sba1 unit, and one without any co-chaperone. As the crystal structure contains truncated loop domains without any connection to the protein bulk which cause the appearance of artificial localized normal modes, we removed amino acids 208 to 267 from the Hsp90 dimer, and amino acids 130 to 135 from Sba1.

Simulation data was analyzed using Numpy \autocite{Oliphant.2007}, Scipy \autocite{Virtanen.2020} and pandas \autocite{McKinney.2015} libraries. Structural data was visualized using Pymol \autocite{PyMOL} and VMD \autocite{Humphrey.1996}.

\section*{Conflict of interest}

The authors declare that there is no conflict of interest.

\section*{Data availability}

Neutron data from IN15 and IN16b are associated with the beamtime \mbox{proposal 8-04-838} \autocite{804838}. SANS data were recorded on D11 during an internal beamtime and are available on request.

\section*{Author contributions}

B.S., C.B, F.R., F.S., T.S. and T.H. designed the research; B.S., C.B, M.G., T.S, I.H. performed the measurements; all authors contributed to the data analysis and interpretation. S. W. performed MD simulations; B.H. and B.S. prepared the samples; all authors wrote the manuscript.

\section*{Acknowledgements}

This work was supported by the European Research Council (grant agreement No. 681891) and the Deutsche Forschungsgemeinschaft
(DFG) under Germany's Excellence Strategy (CIBSS EXC-2189 Project ID 390939984), via grant WO 2410/2-1 within the framework of the Research Unit FOR 5099 ''Reducing complexity of nonequilibrium'' (project No. 431945604), via the Project-ID 403222702 -- SFB 1381 and DFG grant No.~INST 37/935-1 FUGG. The authors acknowledge support by the bwUniCluster computing initiative, the High Performance and Cloud Computing Group at the Zentrum f\"ur Datenverarbeitung of the University of T\"ubingen and the state of Baden-W\"urttemberg through bwHPC. F.S. acknowledges support from the BMBF (FKZ 05K19VTB).

We thank B. Schuler and D. Nettels for insightful comments and discussions and for providing open acces to their software (Fretica). We thank J. Vorreiter and J. Thurn for beamtime support. We thank the PSCM and ESRF (Grenoble) for sharing their laboratory resources. We thank S. Pr\'evost and R. Schweins for data collection at beamline D11 and G. Stock for helpful discussions. We thank J. Vorreiter, J. W\"orner, M. Werner, S. Weber and E. Bartsch for support during sample preparation.


\end{document}


\renewcommand\thefigure{S\arabic{figure}}
    \renewcommand\thetable{S\arabic{table}}
    \renewcommand\theequation{S\arabic{equation}}
        \title{Supporting Information: The onset of molecule-spanning dynamics in a multi-domain protein}
    \author[1]{Benedikt Sohmen}
    \author[2,3]{Christian Beck}
    \author[3]{Tilo Seydel}
    \author[3]{Ingo Hoffmann}
    \author[1]{Bianca Hermann}
    \author[4]{Mark N\"uesch}
    \author[3]{Marco Grimaldo}
    \author[2]{Frank Schreiber$^*$}
    \author[5]{Steffen Wolf$^*$}
    \author[6]{Felix Roosen-Runge$^*$}
    \author[1,7]{Thorsten Hugel$^*$}
    \affil[1]{\normalsize Institute of Physical Chemistry, University of Freiburg, Albertstrasse 21, 79104 Freiburg, Germany}
    \affil[2]{Institute of Applied Physics, University of T\"ubingen,  Auf der Morgenstelle 10, 72076 T\"ubingen, Germany}
    \affil[3]{Institut Max von Laue - Paul Langevin, 71 avenue des Martyrs, 38042 Grenoble, France}
    \affil[4]{Department of Biochemistry, University of Zurich, Winterthurerstrasse 190, CH-8057 Zurich, Switzerland}
    \affil[5]{Biomolecular Dynamics, Institute of Physics, University of Freiburg, Hermann-Herder-Strasse 3, 79104 Freiburg, Germany}
    \affil[6]{Department of Biomedical Sciences and Biofilms-Research Center for Biointerfaces (BRCB), Malm\"o University, 20506 Malm\"o, Sweden}
    \affil[7]{Signalling Research Centers BIOSS and CIBSS, University of Freiburg, Sch\"anzlestrasse 18, 79104 Freiburg, Germany}
    \affil[ ]{$^*$contact details:
        \href{mailto:frank.schreiber@uni-tuebingen.de}{frank.schreiber@uni-tuebingen.de};
        \href{mailto:steffen.wolf@physik.uni-freiburg.de}{steffen.wolf@physik.uni-freiburg.de};
        \href{mailto:felix.roosen-runge@mau.se}{felix.roosen-runge@mau.se};
        \href{mailto:thorsten.hugel@physchem.uni-freiburg.de}{thorsten.hugel@physchem.uni-freiburg.de}}
    \maketitle
	\newpage
\section{Supplementary Figures}

\begin{figure}[h]
	\centering
	\includegraphics[width=0.5\textwidth]{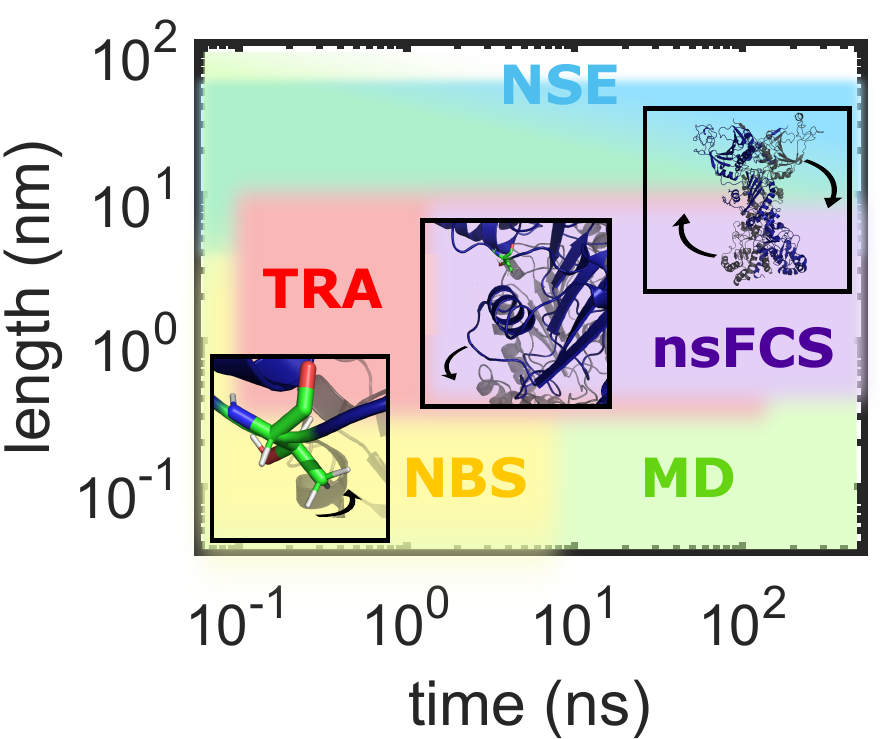}
	\caption{Overview of employed techniques. Nanosecond fluorescence correlation spectroscopy (nsFCS), time-resovled anisotropy (TRA), neutron spin echo (NSE), neutron backscattering (NBS) and full-atom molecular dynamics (MD) simulations cover dynamics from ps to µs and length scales from sub-nm to $\sim$100\,nm.}
	\label{fig:intro-technique-overview}
\end{figure}

\begin{figure}[h]
	\centering
	\includegraphics[width=0.6\textwidth]{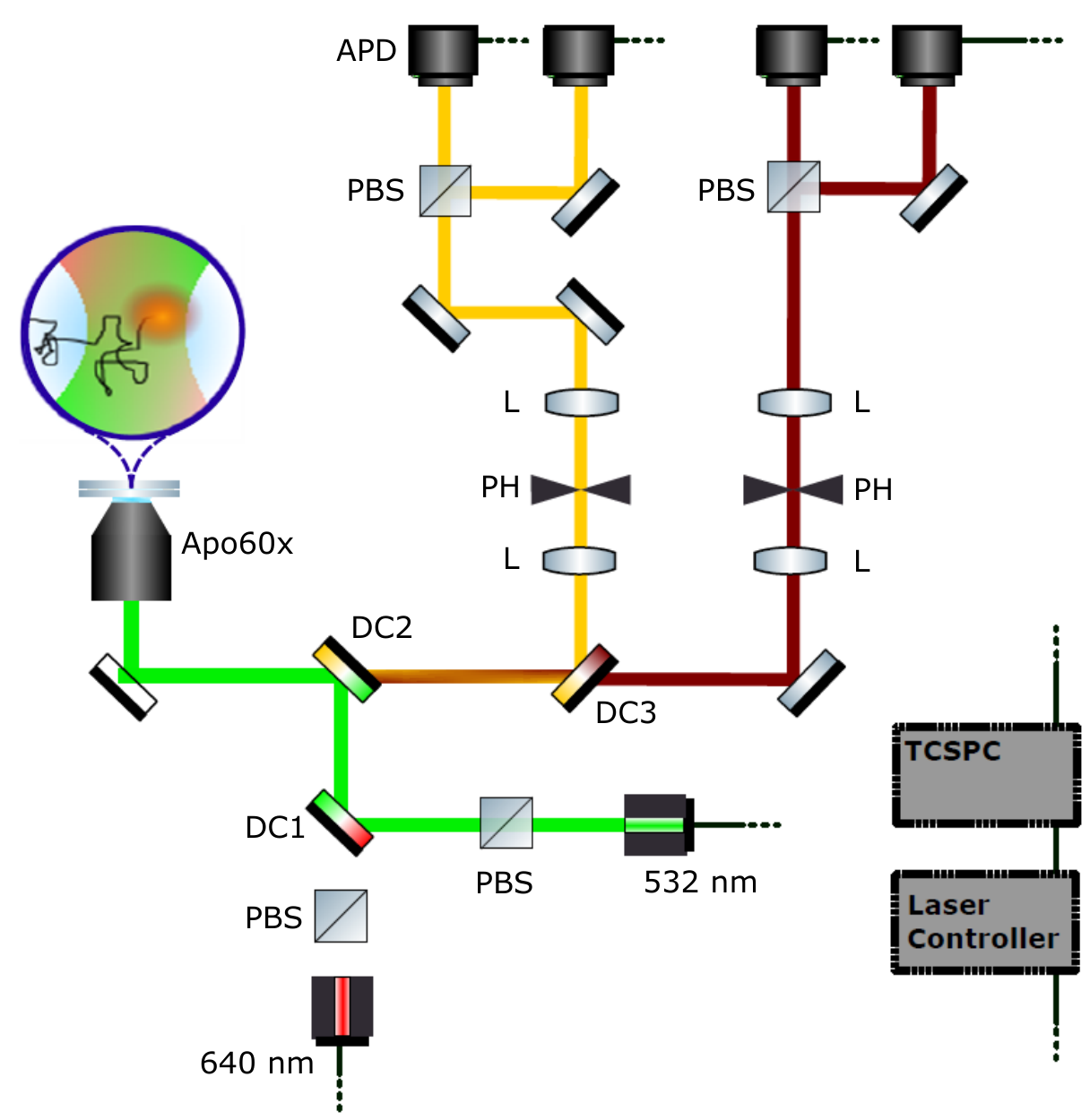}
	\caption{Scheme of the confocal single-molecule fluorescence set-up, here operated in continuous wave (cw) excitation mode. Depending on the type of experiment, dyes were excited with 532\,nm (cw) or 532\,nm and 640\,nm (pulsed-interleaved). Polarizing beam splitters (PBS) were used to generate polarized excitation light and polarization-sensitive detection. Spectral overlay and separation was achieved by dichroic mirrors (DC): overlay of green and red excitation pulses (DC1), separation of excitation light, scattered light and fluorescence emission (DC2) and separation of donor- and FRET-based acceptor emission (DC3). Light was focussed and re-collected by an apochromat (Apo 60x). Lenses (L) in combination with pinholes (PH) were used to achieve confocal single-molecule detection. Single photon detection with picosecond time-resolution was achieved by avalanche photon diodes (APD).}
	\label{SI:fig:SIconfocalsetup}
\end{figure}

\begin{figure}[h]
	\centering
	\includegraphics[width=0.5\textwidth]{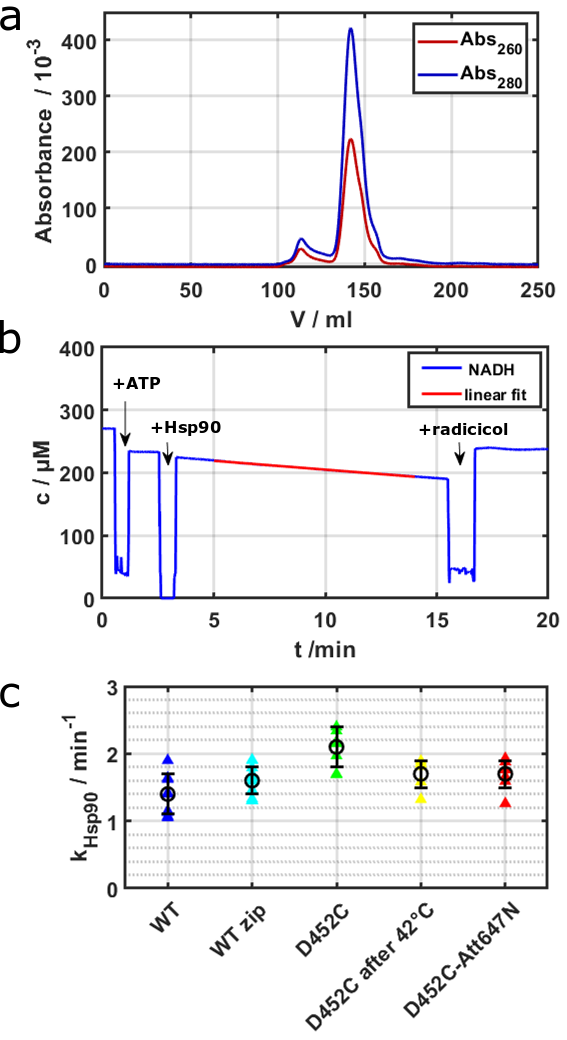}
	\caption{a) Representative SEC profile of the last purification step of yeast Hsp90 wildtype. The chromatogram shows two main fractions at 113\,ml and 142\,ml, respectively. The peak around 142\,ml corresponds to the Hsp90 dimer and was pooled excluding the shoulder fraction. Size exclusion was performed on a Superdex 200 XK 26/600.
		b) ATPase assay of yeast Hsp90 wildtype. The assay couples ATP hydrolyzation to the oxidation of NADH to colorless NAD$^+$ which was monitored over time on a Lambda35 UVvis spectrometer. After addition of 2\,mM ATP and 2\,$\upmu$M Hsp90 a linear decrease of the absorbance at 340\,nm was obtained. From the slope, an ATPase rate of 1.4$\pm$0.3\,min$^{-1}$ is obtained which is consistent with previous studies\autocite{Girstmair.2019}. Final addition of the specific Hsp90 inhibitor radicicol stops NADH consumption which excludes the presence of non-Hsp90 ATPases. c) Summary of ATPase rates measured for yeast Hsp90 wildtype (WT), yeast Hsp90 with a C-terminally inserted coiled-coil motif (WT zip), the cysteine variant D452 and the cysteine variant after the labelling procedure. As control the labelling procedure was performed with and without dye (D452C after 42$^{\circ}$C and D452CAtt647N, respectively). All ATPase assays resulted in similiar ATPase rates which makes Hsp90-related differences between fluorescence and neutron experiments as well as hindered bio-functionility unlikely.}
	\label{SI:fig:SIsecATPase}
\end{figure}

\begin{figure}[h]
	\centering
    \includegraphics[width=0.95\textwidth]{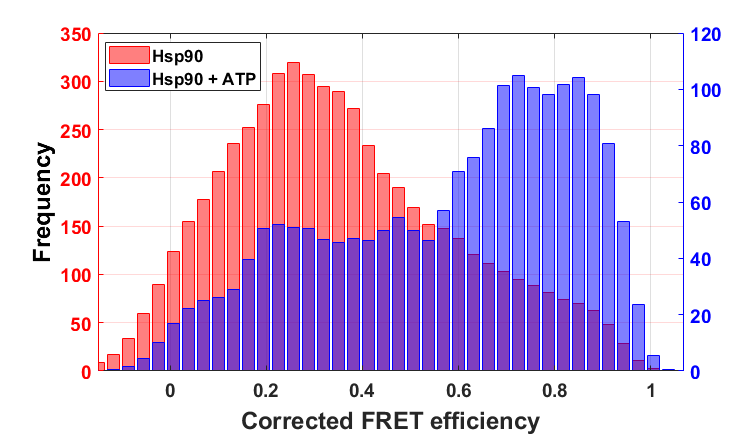}
	\caption{FRET efficiency histogram of Hsp90 in the apo condition (red) and with 2\,mM ATP (blue) in D$_2$O based buffer. Hsp90 was labelled at position 452 with Atto550 and Atto647N as donor and acceptor, respectively. ATP depopulates the open conformational state of Hsp90 (here at E$\sim$0.28). This is the known behaviour in H$_2$O-based buffer and gives evidence that D$_2$O does not have a crucial effect on the biologically relevant open-close dynamics of Hsp90.}
	\label{SI:fig:SIhsp90d2o}
\end{figure}

\clearpage

\begin{figure}[h]
	\centering
	\includegraphics[width=0.95\textwidth]{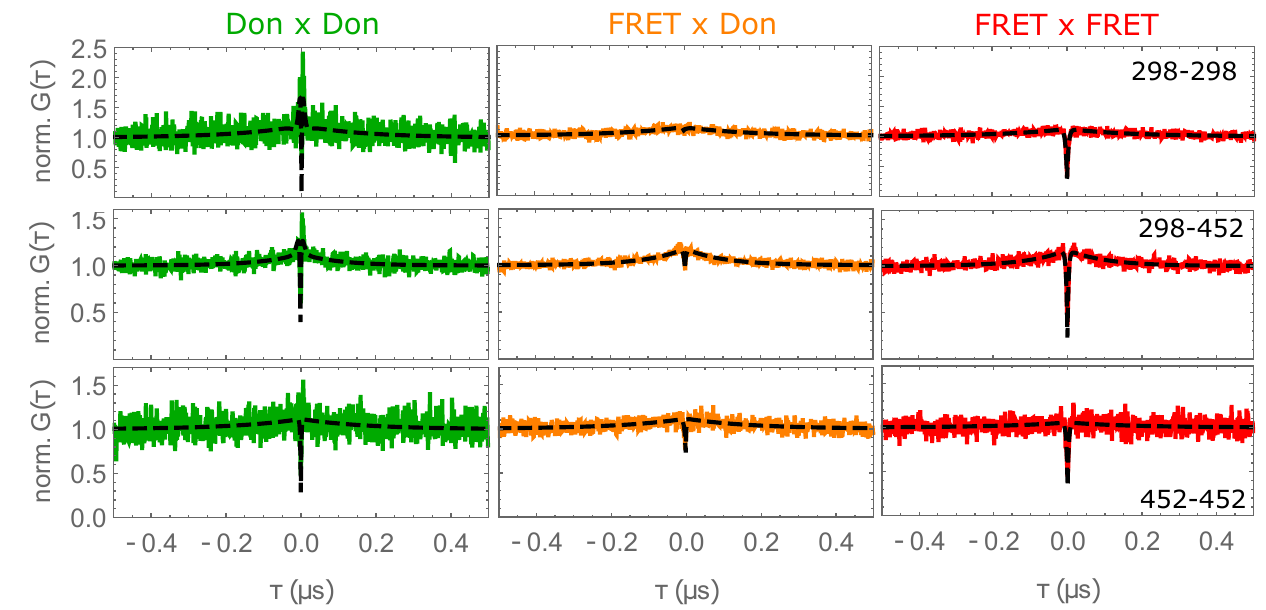}
	\caption{Substate-specific nsFCS data and fits of Hsp90 FRET pairs 298-298, 298-452 and 452-452 with AMPPNP. 
	Closed-state dynamics of Hsp90 were analysed by selecting only single-molecule events with proximity ratios $>$0.3. For each FRET pair Don$\times$Don (green), FRET$\times$Don (orange) and FRET$\times$FRET (red) correlations are shown. Each data set is described by an individual antibunching time ($\sim$3\,ns) and a global bunching time on the $\sim$100\,ns time-scale. For data with donor position 298 involved an additional bunching term was added to account for Trp quenching which resulted on a time-scale of $\sim$5\,ns. For a systematic investigation of Trp quenching please refer to Fig.\,\ref{SI:fig:SInsFCSDyes}. All fit parameters are summarised in Tab. \ref{SI:tab:nsFCShsp90fret}.}
	\label{SI:fig:SInsFCSHsp90subpop}
\end{figure}

\begin{figure}[h]
	\centering
    \includegraphics[width=0.95\textwidth]{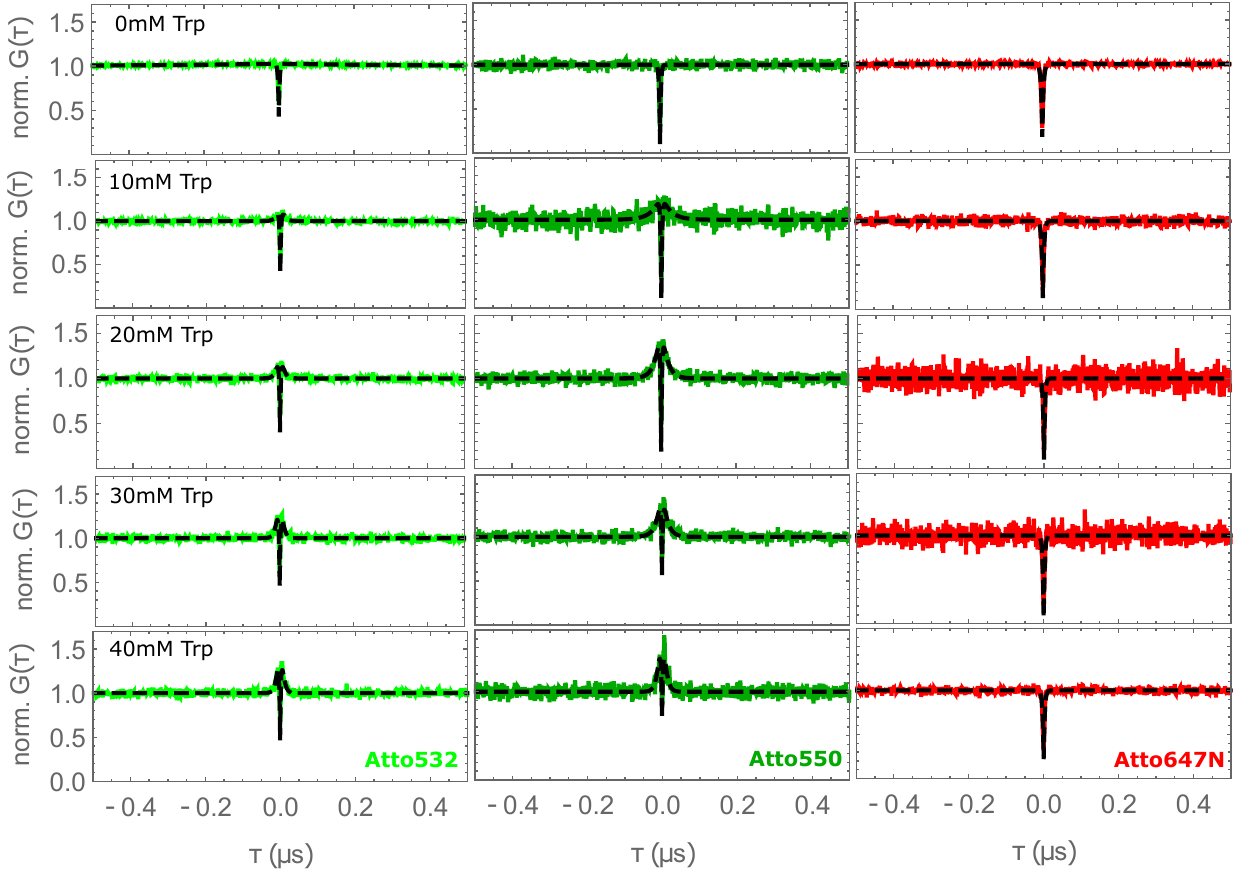}
	\caption{nsFCS data and fits of Atto532, Atto550 and Atto647N with increasing tryptophan concentration from 0-40\,mM. Trp-quenching of Atto532 and Atto550 is revealed by the bunching component which becomes stronger at higher Trp-concentrations. In contrast, no Trp-specific effect was detected for Atto647N. For details on the fit and fit parameters please refer to Tab. \ref{SI:tab:nsFCSdyes}}
	\label{SI:fig:SInsFCSDyes}
\end{figure}

\begin{figure}[h]
	\centering
	\includegraphics[width=0.95\textwidth]{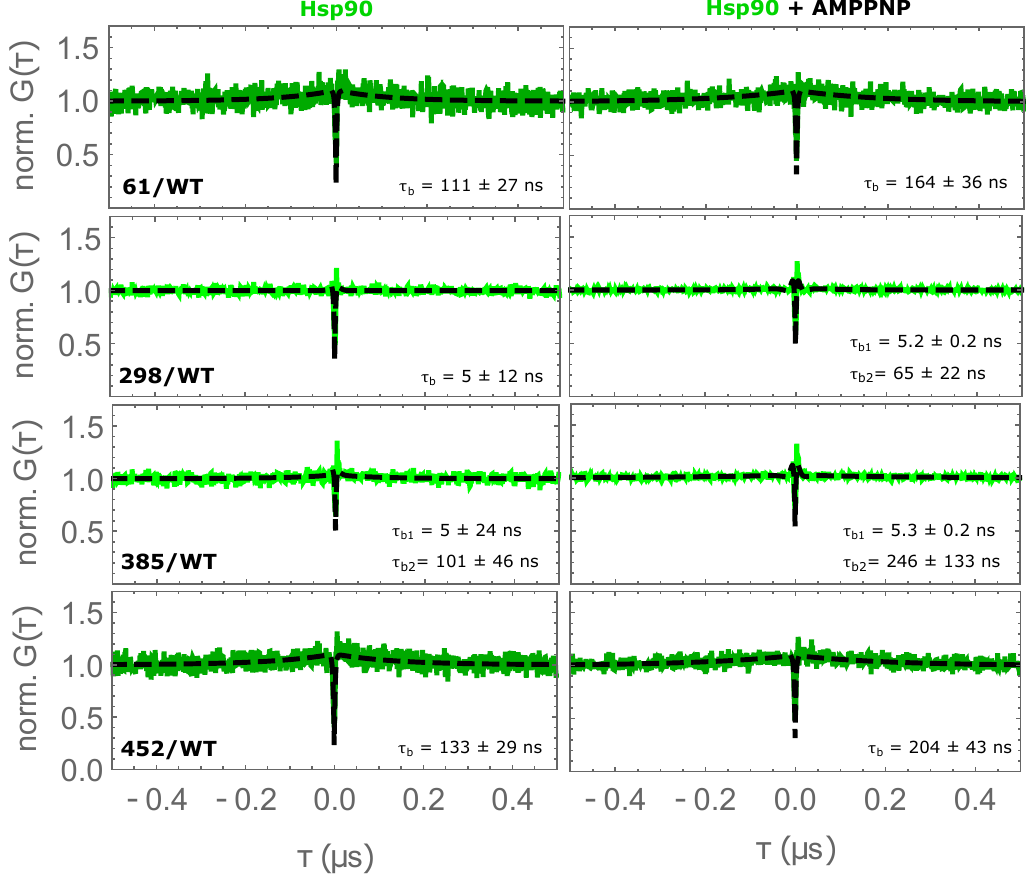}
	\caption{nsFCS data and fits for Hsp90 labeled at positions 61, 298, 385 and 452 with and without AMPPNP. For position 61 and 452 Atto550 and for positions 298 and 385 Atto532 was used as fluorescent label, respectively. Fit models include an antibunching component ($\sim$3\,ns) and a bunching component. At positions 61 and 452 the bunching mode occurred on the $\sim$100\,ns time-scale. Faster bunching times were obtained in absence of AMPPNP, when Hsp90 is mainly in its open state. For the 298 variant, we obtained a bunching component at $\sim$8\,ns associated to Trp quenching. Please refer to Fig.\,\ref{SI:fig:SInsFCSDyes} for the investigation of Trp quenching and to Tab.\,\ref{SI:tab:nsFCShsp90donoronly} for an overview of all fit parameters.}
	\label{SI:fig:SInsFCSHsp90DonorOnly}
\end{figure}

\begin{figure}[h]
	\centering
	\includegraphics[width=0.99\textwidth]{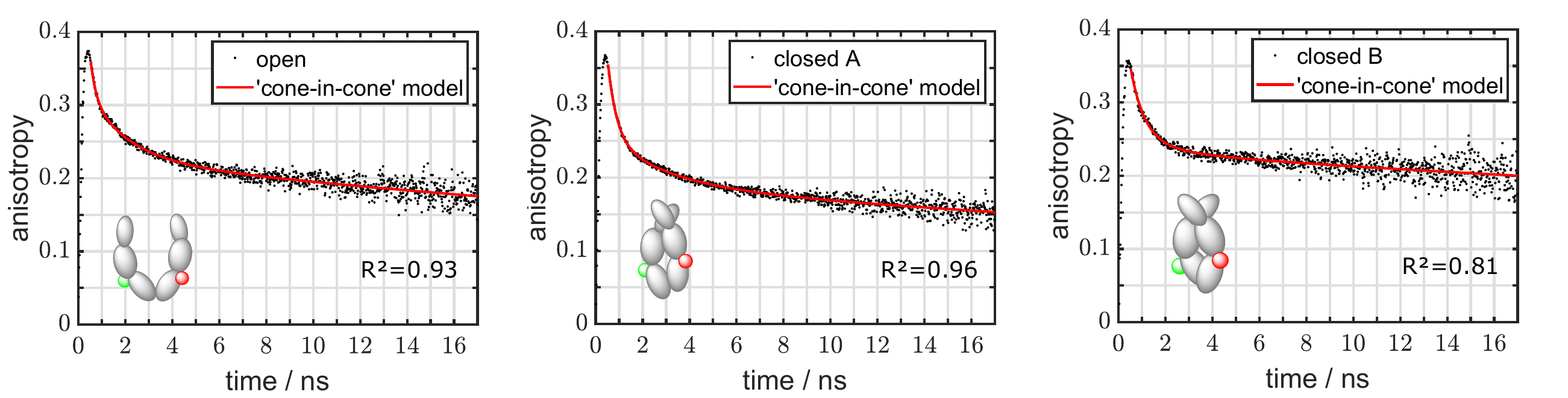}
	\caption{Time-resolved single-molecule acceptor anisotropy decays of Hsp90. The FRET pair 452-452 was labeled with Atto550 and Atto647N and measured in presence of AMPPNP at 21$^{\circ}$C. FRET efficiency vs. stoichiometry analysis was applied to identify conformational sub-states of Hsp90 (open, closed A, closed B). Data were analysed using the \textit{cone-in-cone} model \autocite{Schroder.2005}. The fit results are shown in Tab.\ref{SI:tab:aniso}.}
	\label{SI:fig:SIaniso}
\end{figure}
\clearpage

\begin{figure}[h]
	\centering
	\includegraphics[width=0.99\textwidth]{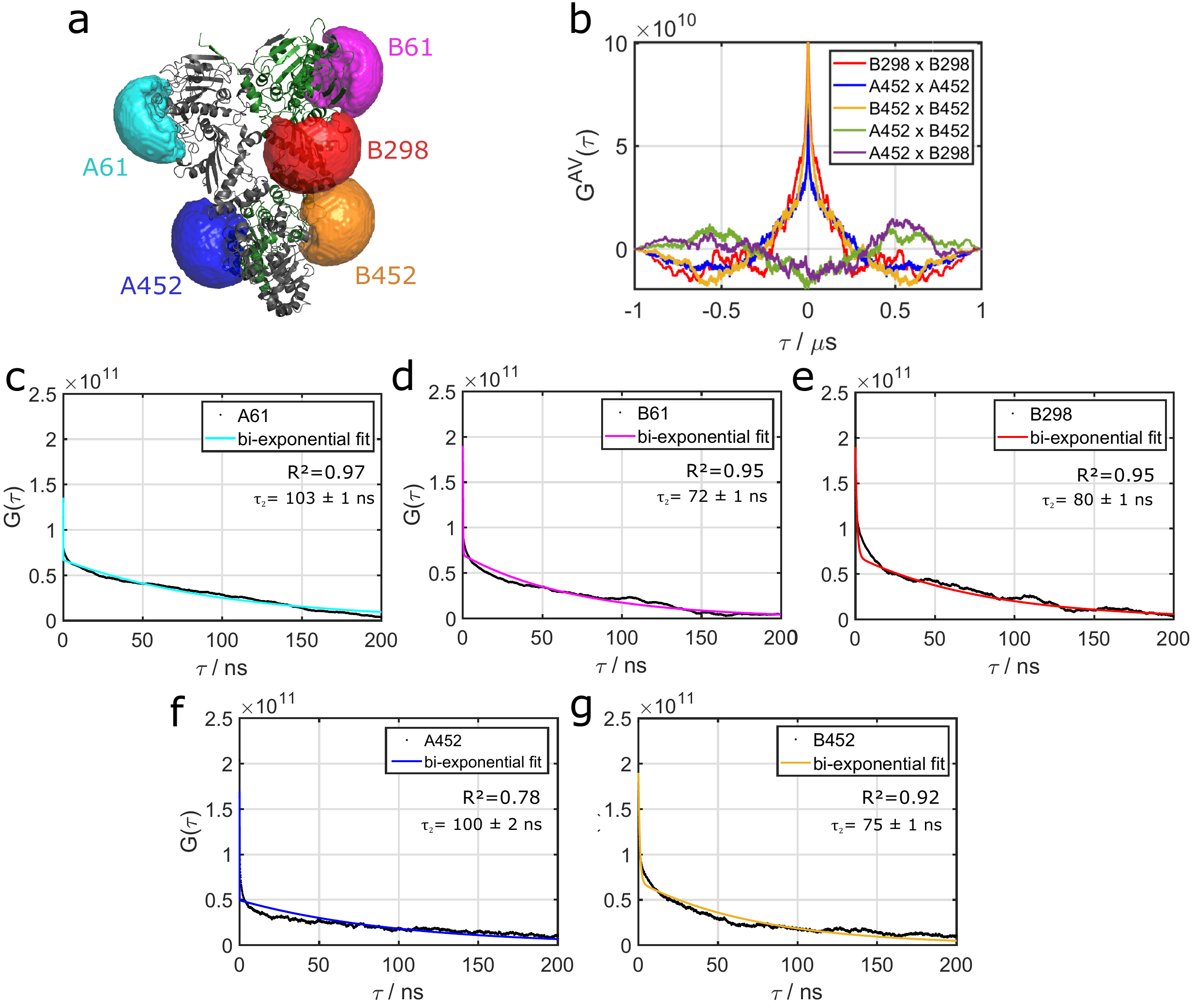}
	\caption{MD-based accessible dye volume correlation analysis for Hsp90 with two AMPPNP molecules. a) Structural MD snapshot of the Hsp90 dimer with accessible dye volumes at positions 61, 298 and 452 at chain A and/or B, respectively. b) Accessible dye volume auto- and cross-correlations. c)-f) Accessible dye volume autocorrelations with unconstrained bi-exponential fits at respective positions in the Hsp90 dimer. Correlations at 70-100\,ns are observed which hint towards local structural dynamics affecting the accessible dye volumes. See Tab.\ref{SI:tab:AVcorr} for all fit results.}
	\label{SI:fig:SIAVcorrfits}
\end{figure}

\begin{figure}[ht]
	\includegraphics[width=0.95\textwidth]{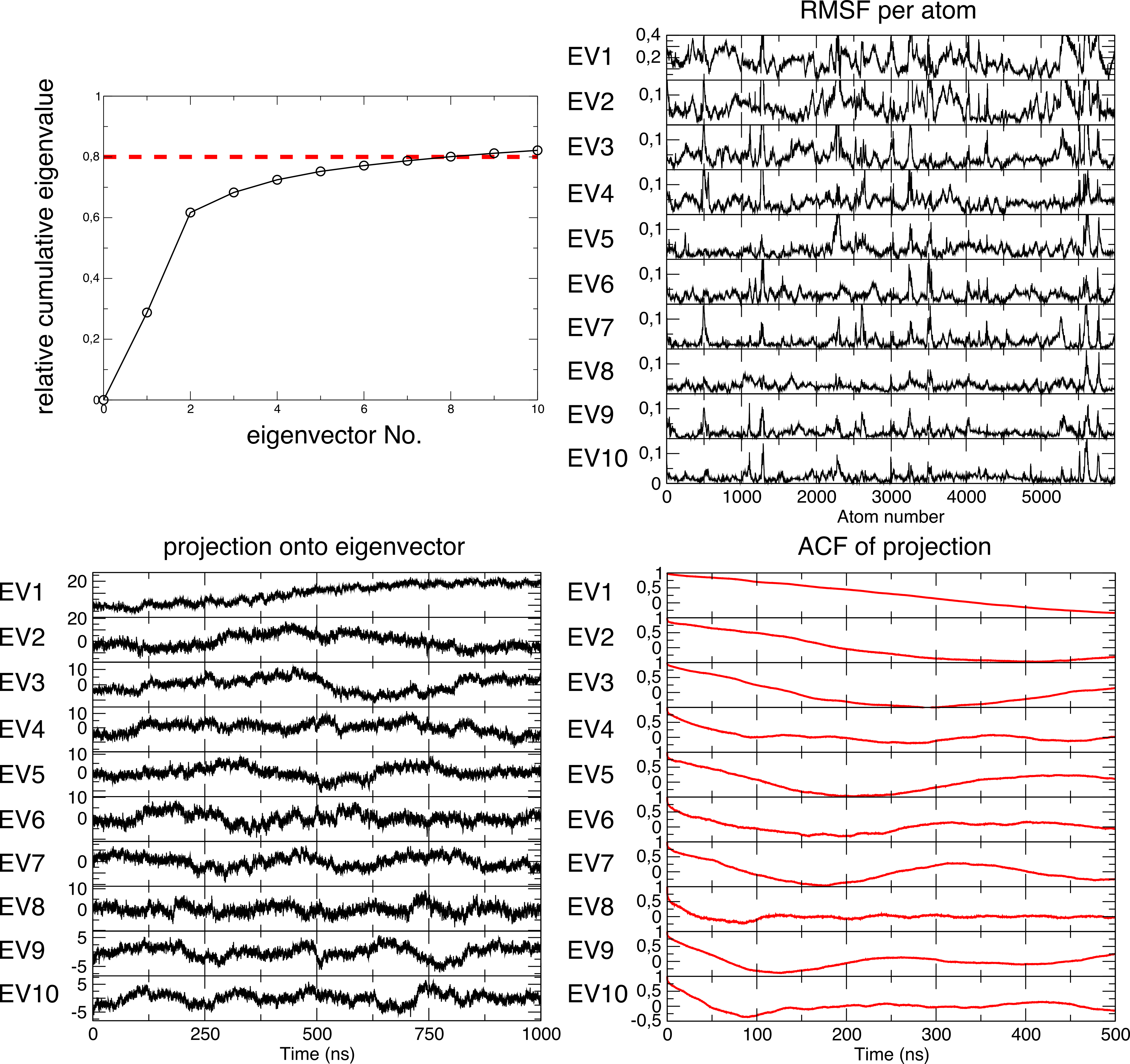}
	\caption{Results from cartesian PCA of one exemplary 1~$\upmu$s simulation. Top left: Cumulative relative eigenvalues. While eigenvectors 1 and 2 contain the majority of variance, eight eigenvectors need to be considered to cover 80\% of the closed dimer's dynamics. Top right: root mean square fluctuations of individual atoms covered in the first ten eigenvectors. All ten vectors represent global motions involving the full dimer. Bottom: projections of one exemplary trajectory onto eigenvectors and their ACF. While all projections seem to contain slow oscillations, they are overlaid by fluctuations on short time scales.}
	\label{SI:fig:cPCA}
\end{figure}

\begin{figure}
	\includegraphics[width=0.95\textwidth]{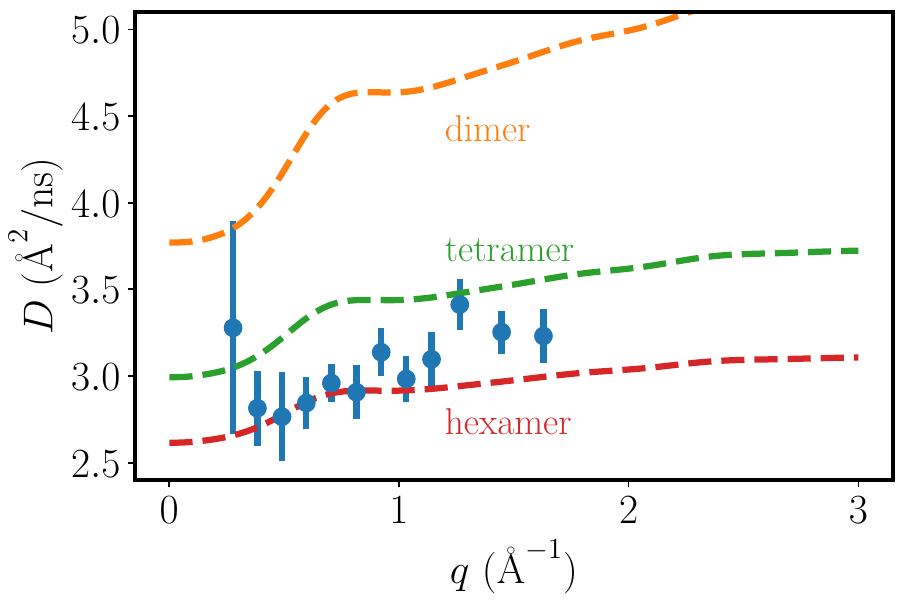}
	\caption{Comparing the rigid-body contribution from pure dimer, tetramer and
		hexamer solutions (dashed lines) with the experimental signature of
		$D_{\textrm{eff}}$ (symbols) suggests the presence of oligomeric states. Given
		that the overall $q$ signature seems conserved for the different
		oligomers, we do not use an assumed polydisperse system in our modeling,
		but restrict ourselves to a pure hexamer solution.}
	\label{fig:nse-si}
\end{figure}

\begin{figure}\begin{center}
		\begin{subfigure}[t]{.7\textwidth}
			\includegraphics[width=\textwidth]{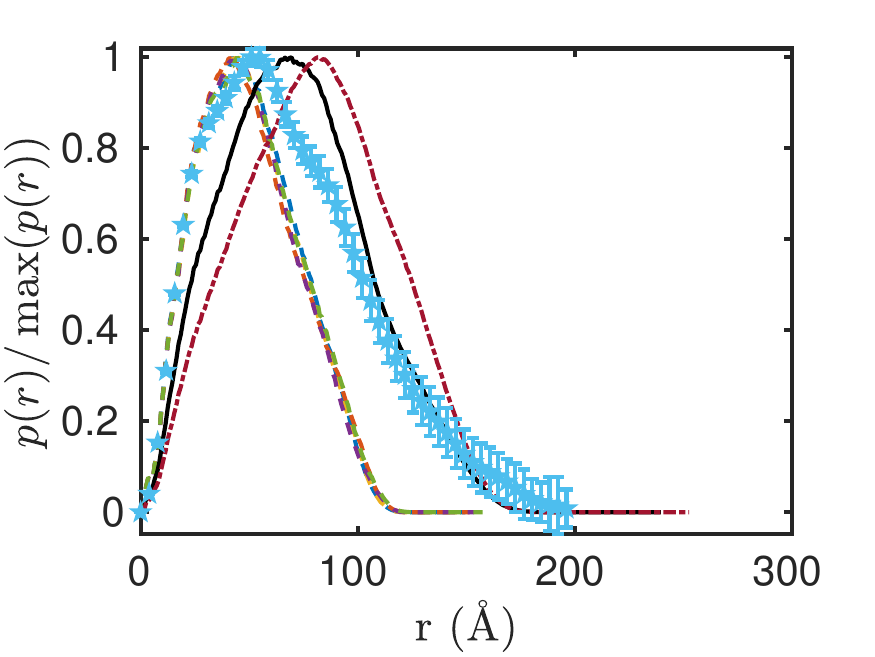}
			\caption{Comparison of $p(r)$ obtained from SANS data (light blue stars and error bars), five independent 1\,$\upmu$s MD trajectories (dashed lines) and thereon-based tetramers (solid line) and hexamers (dashed dotted line).}
			\label{pr:Sim}
		\end{subfigure}
		\\
		\begin{subfigure}[t]{.7\textwidth}
			\includegraphics[width=\textwidth]{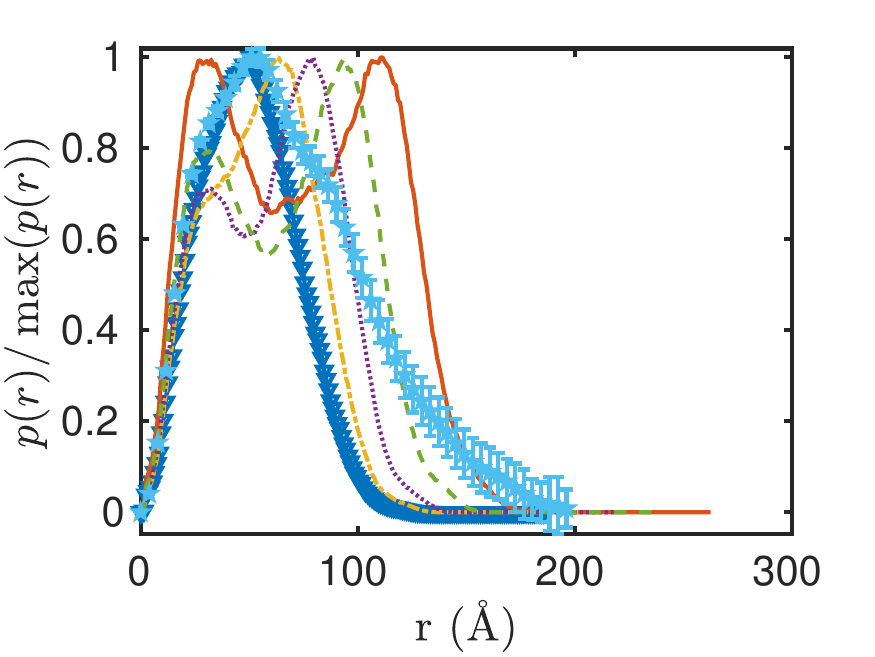}
			\caption{Comparison of $p(r)$ obtained from SANS data (light blue stars and error bars) and different quaternary structures evolving from a closed (solid line with triangles) via different sub-states (dashed-dotted, dotted, dashed) to an open state (solid line).}
			\label{pr:quat}
		\end{subfigure}
	\end{center}
	\caption{Pair distribution functions $p(r)$ of Hsp90: Stars and error bars (light blue) represent $p(r)$ determined from SANS data with SASView. Additional lines correspond to calculated pair distribution functions based on MD structures as specified in the subcaptions. All curves are rescaled in such a way that max($p(r)$)=1.}
	\label{pr}
\end{figure}

\begin{figure}
    \begin{center}
    \begin{subfigure}[t]{0.75\textwidth}
        \includegraphics[width=\linewidth]{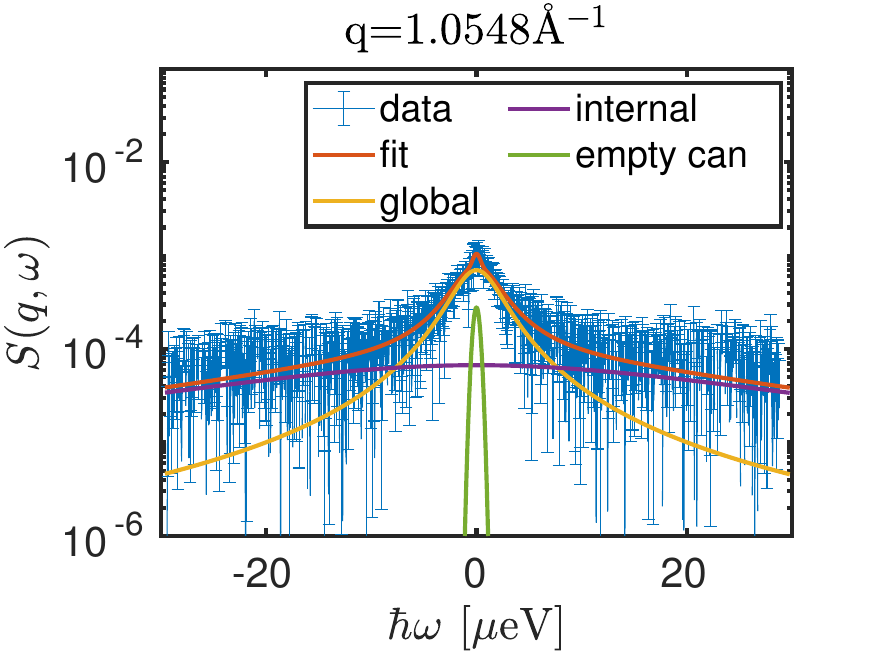}
        \caption{NBS fit at $q=1\,\textrm{\AA}^{-1}$. The internal and global diffusive contributions as well as the contribution of the empty can are shown in violet, yellow and green, respectively.}
        \label{fig:QENSFIT}
    \end{subfigure}\\
\begin{subfigure}[t]{0.75\textwidth}
\includegraphics[width=\linewidth]{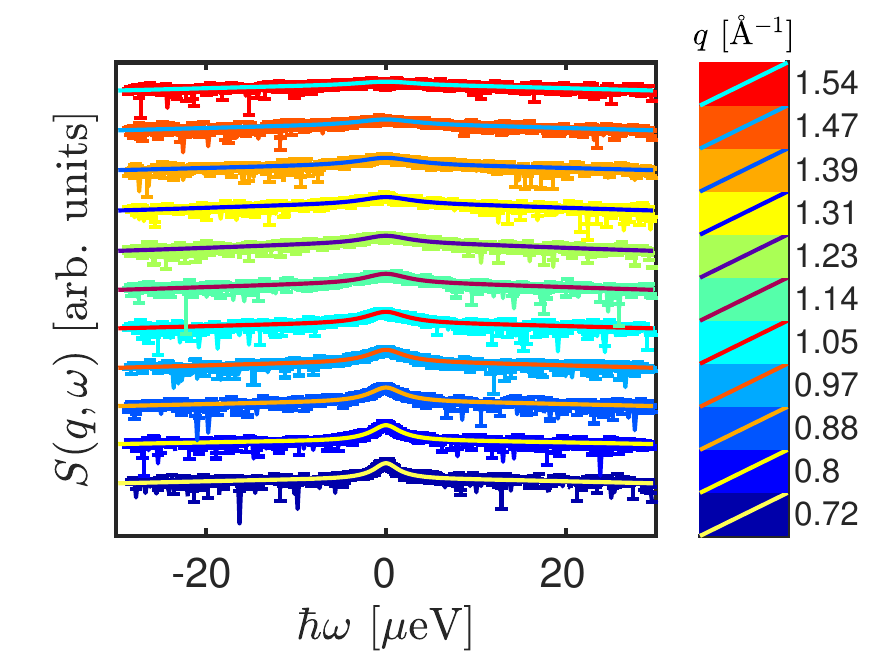}
\caption{NBS fit for different $q$-values. The total fit is shown as solid line. For each $q$ value, the data was binned and shifted by a factor of 1000 for better visibility. The $q$ values and corresponding fits are color-coded in the legend on the right. A clear broadening with $q$ is visible.}
\label{SI:fig:QENS}
\end{subfigure}
\end{center}
	\caption{Fits of the neutron backscattering spectra of Hsp90 with 2\,mM AMPPNP.}
\end{figure}

\begin{figure}
    \centering
    \includegraphics[width=.7\textwidth]{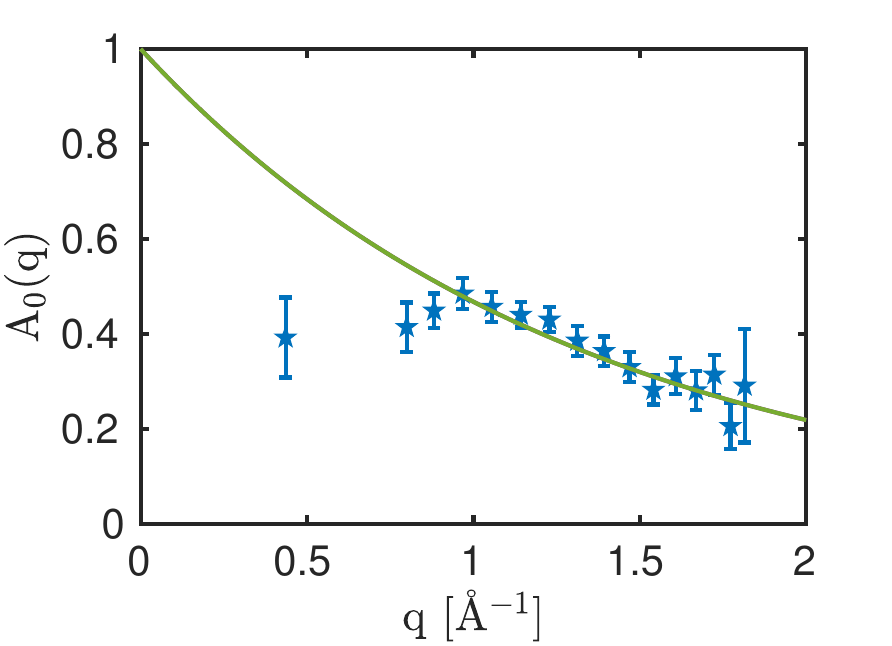}
    \caption{Elastic incoherent structure factor (EISF) obtained from the NBS analysis with corresponding fit (Eq.\,\ref{eq:EISF}).}
    \label{SI:fig:EISF}
\end{figure}
\newpage
    \thispagestyle{empty}
\begin{figure}[ht]
	\includegraphics[width=0.95\textwidth]{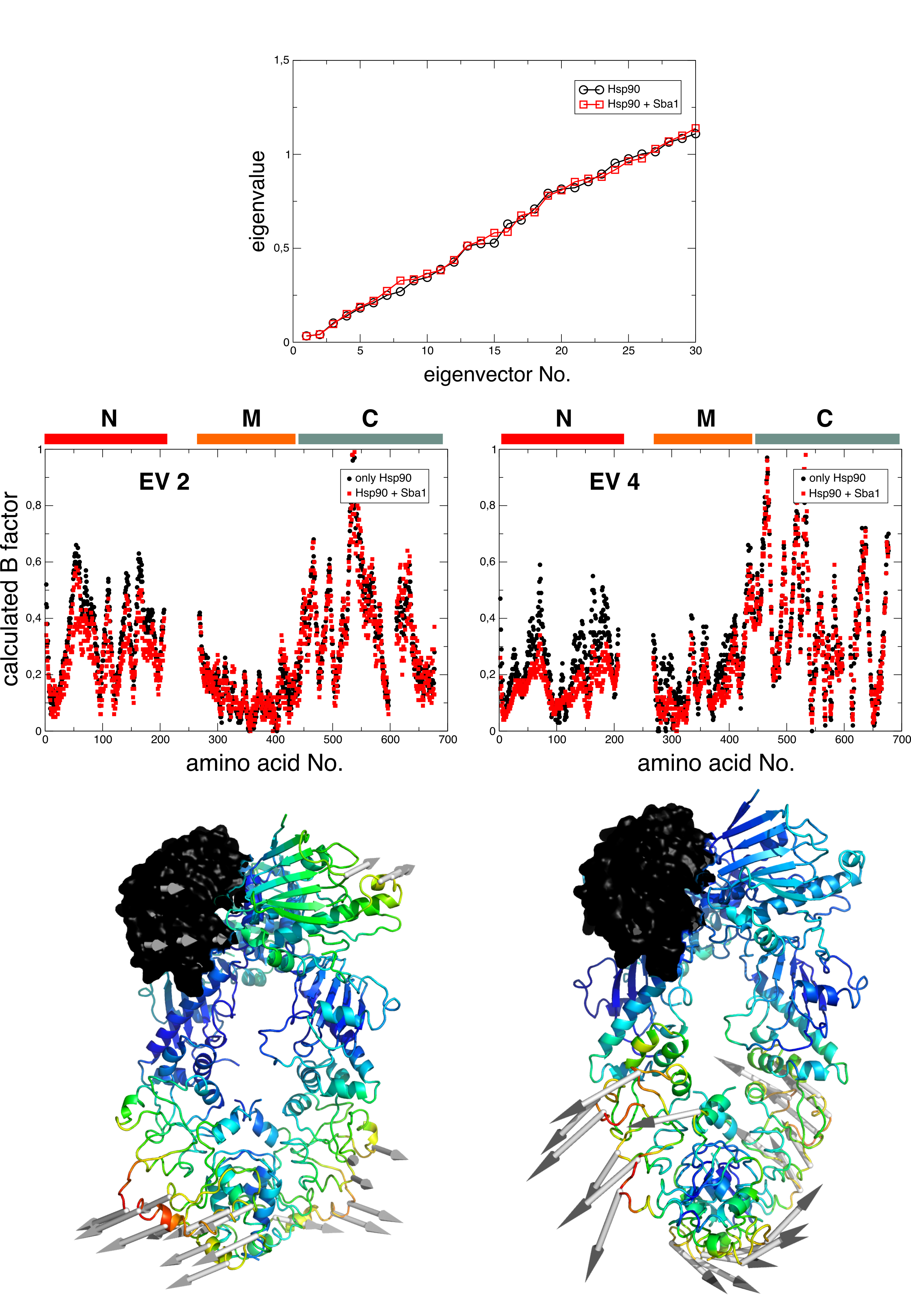}
	\caption{Anisotropic network model analysis. Top: eigenvalues of eigenvectors ordered according to lowest value / slowest oscillation (trivial first six modes not shown). Differences between Hsp90 with and without Sba1 are negligible. Middle: B factors per C$_{\alpha}$ atom calculated for exemplary normal modes 2 and 4. While mode 2 does not exhibit any significant differences between Hsp90 with and without Sba1, mode 4 exhibits clearly reduced dynamics around the N terminal domain. Bottom: Hsp90-Sba1 complex dynamics for normal modes 2 and 4, respectively. Sba1 as black surface. Hsp90 colored according to B factor with high factors in red, low factors in blue. Arrows indicate atoms with the highest contribution to / largest motion within the displayed normal modes.}
	\label{SI:fig:Sba1_ANM}
\end{figure}

\clearpage

\section{Legends of Supplementary Movies}

\begin{itemize}
	\item\textbf{Supplementary Movie 1:} Morphing along the first eigenvector from the cartesian PCA.  The displayed motion represents a morphing between the two structures appearing during simulations with minimal and maximal dot product of the cartesian coordinates of the protein and the first eigenvector. Hsp90 dimer displayed as cartoon representation.
	
	\item\textbf{Supplementary Movie 2:} Morphing along the second eigenvector from cartesian PCA. The displayed motion represents a morphing between the two structures appearing during simulations with minimal and maximal dot product of the cartesian coordinates of the protein and the second eigenvector. Hsp90 dimer displayed as cartoon representation.
	
	\item\textbf{Supplementary Movie 3:} Motion along the first eigenvector from cartesian PCA during one microsecond MD trajectory.  The component of motion along the first eigenvector represents a fluctuating twist motion of the protein. Hsp90 dimer displayed as cartoon representation.
	
	\item\textbf{Supplementary Movie 4:} Motion along the second eigenvector from cartesian PCA during one microsecond MD trajectory.  The component of motion along the second eigenvector represents a fluctuating twist motion of the protein. Hsp90 dimer displayed as cartoon representation.
\end{itemize}

\clearpage

\section{Supplementary Methods}

\subsection{Sample Preparation}

Gene expression and subsequent protein purification was performed as previously described \autocite{Schmid.2020}. pET derived expression plasmids contained the \textit{hsp82} gene from baker's yeast with a cleavable N-terminal His$_6$-SUMO-tag. For fluorescent labelling a cysteine was introduced at position 61, 298 and 452 via site-directed mutagenesis, respectively. \textit{E. coli} BL21 Star (DE3) cells were transformed with the respective plasmid and cultivated in \textit{lysogeny broth} medium supplemented with kanamycin at 37\,$^{\circ}$C. Expression was induced with 1\,mM IPTG at OD$_{600}$ between 0.6-0.8. Cells were harvested 4\,h after induction and stored at -20\,$^{\circ}$C.

All buffers contained 40\,mM HEPES and 150\,mM NaCl at pH 7.5. Cleared lysate of His$_{6}$-SUMO-tagged yeast Hsp90 WT or D452C was applied to a HisTrap HP column (Cytiva, 10\,ml) performing affinity chromatography (AC) and eluted with a gradient from 20\,mM to 1\,M imidazole in the buffer. For ion exchange chromatography (IEX), the salt concentration was adjusted to 30\,mM NaCl for binding and 1\,M for elution.

His$_6$-SUMO-tags were cleaved in presence of SenP protease while dialyzing against imidazole free buffer over night. A second HisTrap was applied to remove uncleaved fusion proteins and tags, leaving native like Hsp90 in the flow through. The flow through was diluted to 30\,mM NaCl and applied to a HiTrap Q column (Cytiva, 10\,ml). After elution with a NaCl gradient, target proteins were polished by a final size exclusion chromatography (SEC) step (S200, XK 26/600), concentrated and flash frozen in liquid nitrogen.

\subsection{Nanosecond fluorescence correlation spectroscopy}

For a clear data interpretation we investigated the contributions of the used fluorescent labels to the nanosecond range of an FCS curve. To this end we measured freely diffusing Atto532, Atto550 and Atto647N in buffer (Fig. S5, top row). Atto532 and Atto550 were excited with green continuous wave (cw) illumination (84\,$\upmu$W and 287\,$\upmu$W, respectively) and Atto647N with red cw illumination (138\,$\upmu$W). The obtained signals were correlated whereby `Don’ denotes the donor signal after donor excitation and `Acc’ the acceptor signal upon acceptor excitation.
The data shows a dip at the minimum lag time which is the typical feature of fluorophore antibunching. A fit model with only one antibuching term (Eq.\,\ref{eq:nsFCSmodelSI}a) delivers a correlation time on the $\sim$3\,ns time scale which is consistent with the fluorescence lifetime of the dyes stated by the manufacturer.

As biomolecules contain aromatic amino acids known to quench fluorescence the interaction between Atto532, Atto550 and Atto647N and tryptophan (Trp) was investigated by measuring a Trp concentration series (10, 20, 30, 40\,mM Trp) for each of the dyes. We find that Trp affects donor and acceptor dyes very differently. In case of Atto647N nsFCS detects no difference between the absence and presence of Trp (see Fig. S5). In contrast, for the donor dyes (Atto532, Atto550), the presence of Trp causes a bump in the low nanosecond range which is more pronounced the higher the Trp concentration is. By extending the fit model with an additional bunching term (Eq.\,\ref{eq:nsFCSmodelSI}b) we obtain increasing bunching weights with increasing Trp concentrations for both, Atto532 and Atto550 (see Tab.S3). However, the bunching times behave differently: While for Atto532 we obtain a Trp-independent correlation time of on average 5.4$\pm$0.3\,ns, for Atto550 the bunching time decreases with increasing Trp concentrations from 15$\pm$4\,ns to 8$\pm$3\,ns. From this we conclude that Atto532 is mainly quenched by Trp collisions while Atto550 additionally exhibits Trp-complex formation.

These findings well explain the $\sim$5\,ns-correlation component observed for Hsp90 experiments which involve position 298 (see Tab. S1 and S2): Position 298 is close to the biologically relevant switch point Trp300 \autocite{Rutz.2018}. Hence, for data with Hsp90 position 298 involved we included an additional bunching term in the Don$\times$Don correlation to consider Trp quenching kinetics (Eq.\,\ref{eq:nsFCSmodelSI}c). Consistent with our interpretation, for FRET pair 298-298 the obtained bunching weight of the Trp-quenching component $c_{b2}$ is higher (6.4$\pm$0.8) than the one of FRET pair 298-452 (0.8$\pm$0.4). Hsp90 nsFCS data sets without a Trp close to the labeling site were well described by Eq.\,\ref{eq:nsFCSmodelSI}b.

In Hsp90, a global bunching time on the $\sim$100\,ns component was obtained by simultaneously optimising the fit to all three correlations: Don$\times$Don, FRET$\times$FRET and Don$\times$FRET. Here, `FRET’ represents the acceptor signal upon donor excitation. The autocorrelations Don$\times$Don and FRET$\times$FRET are always correlations between the perpendicular and parallel part of the signal which removes artefacts from detector afterpulsing. Note that the $\sim$100\,ns component was not caused by triplet kinetics because they occur on the low microsecond time scale \autocite{Wolf.2021} and we chose the excitation powers sufficiently low.

All models used to fit the nsFCS data are summarised in Eq.\,\ref{eq:nsFCSmodelSI}:
\begin{subequations}\label{eq:nsFCSmodelSI}
	\begin{alignat}{2}
		G(\tau) &= a(1-c_{ab}e^{-(\tau-\tau_0)/\tau_{ab}}) \\
		G(\tau) &= a(1-c_{ab}e^{-(\tau-\tau_0)/\tau_{ab}}) (1+c_{b1}e^{-(\tau-\tau_0)/\tau_{b1}}) \\
		G(\tau) &= a(1-c_{ab}e^{-(\tau-\tau_0)/\tau_{ab}}) (1+c_{b1}e^{-(\tau-\tau_0)/\tau_{b1}}) (1+c_{b2}e^{-(\tau-\tau_0)/\tau_{b2}})
	\end{alignat}
\end{subequations}
The models include a scaling factor $a$, the weight of the antibunching mode $c_{ab}$, the weights of up to two bunching modes $c_{b1}$ and $c_{b2}$, the antibunching time $\tau_{ab}$ and the bunching times $\tau_{b1}$ and $\tau_{b2}$, respectively. $\tau_0$ corrects for a small delay of the detection channels.

\subsection{Neutron Scattering}

\subsubsection{Neutron Backscattering}

QENS data were reduced with Mantid \autocite{Arnold.2014} and further analyzed with Matlab 2020b (The MathWorks, Inc.) employing the Optimization and Curve Fitting Toolboxes.

Vanadium measurements were used to determine the resolution function and approximated by two Gaussian functions for each momentum transfer $\hbar q$.

To reduce possible cross-talking between different contributions in the fit, the solvent contribution (D$_2$O) to the scattering function was rescaled to account for the volume excluded by the proteins and subtracted from the scattering signal of the samples \autocite{Beck2018}. In addition, a global fit was performed fixing directly the $q$-dependence of the center-of-mass and internal diffusion linewidths (see below) parameters, thus reducing the number of optimization parameters.

This approach remedied the relatively low protein signal and while imposing more prior knowledge compared to other studies \autocite{Beck2018, Grimaldo.2019a}. The chosen approach nevertheless appears justified, since numerous previous studies have confirmed that the center-of-masss diffusion is of the Fickian type \autocite{Grimaldo_2019_QuartRevBiophys}.
The scattering signal was described using two Lorentzian functions. The width of the first Lorentzian function represents the apparent short-time self diffusion and can be described by a Fickian diffusive process \autocite{Grimaldo_2019_QuartRevBiophys} $\gamma=D_{\textrm{app}}q^2$ while the second Lorentzian function by averaging over all internal diffusive processes in the proteins is described by a $q$-independent width $\Gamma$\comment{add References here}. The contribution of the empty can was modeled with an elastic contribution. The incoherent scattering function thus can be written as
\begin{eqnarray}
S(q,\omega)&=&\mathscr{R}(q,\omega)\otimes\left[\beta_{EC}\delta(\omega)+\right.\notag \\
&&\left.\beta\left(A_0\mathscr{L}_\gamma(\omega)+(1-A_0)\mathscr{L}_{\gamma+\Gamma}(\omega)\right)\right] \label{eq:QENSmodel}
\end{eqnarray}
with $A_0$ being the $q$-dependent elastic incoherent structure factor (EISF) and $\beta_{EC}, \beta$ being $q$-dependent scaling parameters.
The EISF indirectly contains information on the local order within the protein via the geometry of the diffusive motion by which it is determined \autocite{Grimaldo_2019_QuartRevBiophys}, \autocite{volino1980neutron}

Figure \ref{fig:QENSFIT} shows the fit result of Equation \ref{eq:QENSmodel} at $q=1\,\textrm{\AA}^{-1}$. The other $q$ values are depicted in the SI in Figure \ref{SI:fig:QENS}.

The short-time apparent diffusion coefficient $D_{\textrm{app}}$, being a combination of global short-time translational diffusion as well as global short-time rotational diffusion \autocite{Grimaldo_2019_QuartRevBiophys} was determined to a value of $D_{\textrm{app}}=(3.27\pm0.18)~\frac{\textrm{\AA}^2}{\textrm{ns}}$. The width $\Gamma$ of the Lorentzian function describing the internal diffusion corresponds to a characteristic time $\tau=\frac{\hbar}{\Gamma}=\left(23.7\pm2.6\right)\textrm{ps}$. The EISF can be approximated by a diffusion around an equilibrium position of a potential with a Gaussian shape and an effective radius $a$ \autocite{Volino2006}
\begin{equation}
A_0(q)=\exp\left(-\frac{(a\cdot q)^2}{5}\right)\label{eq:EISF}
\end{equation}
with $a=(1.95\pm0.07)~\textrm{\AA}$. The corresponding fit is shown in Figure \ref{SI:fig:EISF}.

\subsection{Pair Distance Distribution Functions}
SANS measurements of a sample with $c_p=10~\textrm{mg/ml}$ in presence of AMPPNP were used to determine the pair distance distribution function $p(r)$. After a constant background subtraction, $p(r)$ was determined via the inversion algorithm of SASView\autocite{SASView}. The results with corresponding error bars are shown in Figure \ref{pr} (light blue stars).
Additionally, pair distribution functions were calculated based on different Hsp90 structures using CaPP 3.12 \autocite{CaPP}.
$p(r)$ was calculated from 1\,$\upmu$s MD-trajectories of five independent runs, for pdb-based tetramers and hexamers (Fig.\,\ref{pr:Sim}) and for different open Hsp90 structures (Fig.\,\ref{pr:quat}) obtained by smFRET-restrained modelling \autocite{Hellenkamp.2017}.

We find that the closed Hsp90 dimers alone are insufficient for a full description of our data. Especially at large distances the data deviates from $p(r)$ calculated based on the closed Hsp90 dimer structure. Therefore, the most likely explanation of our data is that some multimeric species is present in addition to Hsp90 dimers. In addition, likely a certain amount of open states is present, but this is below 10\% in the presence of AMPPNP \autocite{Wolf.2021}. However, we are convinced that our first principle components show eigenmodes of the dimer, because the dimer is still the prevalent species and specific strong new modes from dimers of dimers or trimers of dimers are very unlikely.

\clearpage

\begin{sidewaystable}[ht!]
    \section{Supplementary Tables}
	\subsection*{Summary of nsFCS fit results for FRET-labeled Hsp90}
	\caption{\baselineskip4mm Summary of nsFCS fit results for all measured Hsp90 FRET pairs in presence of AMPPNP. Label positions and used fluorophores are given in the order `donor-acceptor'. `Correlation' specifies the detection channels which were correlated to obtain the correlation function G($\tau$). DxD and AxA stand for the cross-correlations of the parallel and perpendicular donor and acceptor channels, respectively. Cross-correlation of different polarization channels removed artefacts resulting from detector afterpulsing. AxD stands for the cross-correlation of the red and green detection channel. The used fit models are indicated by the capital letters (A, B, C) and differ in the number of bunching components used. Fit parameters are a scaling factor $a$, the weight of the antibunching mode $c_{ab}$, the weights of the bunching modes $c_{b1}$ and $c_{b2}$, antibunching time $\tau_{ab}$, bunching times $\tau_{b1}$ and $\tau_{b2}$, respectively. $\tau_0$ corrects for a small time delay between the detection channels.}
	\centering
	\begin{tabular}{@{} cccccccccccc @{}}
		\hline
		position & fluorophore & correlation & model & a & c$_{ab}$ & c$_{b1}$ & c$_{b2}$ & $\tau_{ab}$  & $\tau_{b1}$ & $\tau_{b2}$ &  $\tau_0$\\
		   &  &  &  &  &  &  &  & / ns & / ns & / ns & / ns \\

		\hline \hline

		298-298 & 550-647N & DxD & C & 1488$\pm$14 & 0.996$\pm$0.012 & 0.20$\pm$0.017 & 6.4$\pm$0.8   & 7.8$\pm$1.2  & 165$\pm$26 & 4.1$\pm$0.2  & -0.49$\pm$0.07\\

		        &          & AxA & C & 1189$\pm$9  & 0.76$\pm$0.08   & 0.125$\pm$0.017 &              & 3.9$\pm$0.6  & 165$\pm$26 &              & -0.09$\pm$0.3\\

		        &          & AxD & C &  885$\pm$6  & 0.05$\pm$0.03   & 0.142$\pm$0.011 &              &              & 165$\pm$26 &              & -2$\pm$3\\

		\hline

		298-452 & 532-647N & DxD & C & 1126$\pm$2  & 0.80$\pm$0.05   & 0.126$\pm$0.008 & 0.8$\pm$0.4  & 2.9$\pm$0.8  & 102$\pm$5  & 5.6$\pm$1.0  & -1.1$\pm$0.07\\

			    &          & AxA & C & 1118$\pm$3  & 0.82$\pm$0.03   & 0.174$\pm$0.007 &              & 3.1$\pm$0.2  & 102$\pm$5  &              & -0.28$\pm$0.09 \\

			    &          & AxD & C &  769$\pm$1  & 0.19$\pm$0.02   & 0.175$\pm$0.005 &              &              & 102$\pm$5  &              &  -1.7$\pm$0.4 \\

	   \hline

		452-452 & 532-647N & DxD & B & 959$\pm$8  & 0.77$\pm$0.11   &  0.12$\pm$0.012  &              & 1.4$\pm$0.3  & 174$\pm$33 &              & -0.6$\pm$0.13\\

				&          & AxA & B & 975$\pm$6  & 0.71$\pm$0.07   & 0.06$\pm$0.012  &               & 3.1$\pm$0.4  & 174$\pm$33 &              & -0.4$\pm$0.2 \\

				&          & AxD & B &  650$\pm$5 & 0.36$\pm$0.05   & 0.115$\pm$0.009  &              &              & 174$\pm$33 &              &  -1.5$\pm$0.3\\

		\hline
	\end{tabular}
	\label{SI:tab:nsFCShsp90fret}
\end{sidewaystable}

\begin{sidewaystable}[p!]
	\subsection*{Summary of nsFCS fit results for singly labeled Hsp90}
	\centering
	\caption{\baselineskip4mm nsFCS analysis of singly-labelled Hsp90 with and without AMPPNP. Atto532 or Atto550 were used as labels at the specified positions. Given fit results of the DxD correlation are a scaling factor $a$, the weight of the antibunching mode $c_{ab}$, the weights of the bunching modes $c_{b1}$ and $c_{b2}$, the antibunching time $\tau_{ab}$, and the bunching times $\tau_{b1}$ and $\tau_{b2}$, respectively. $\tau_0$ corrects for a small time delay between the detection channels.}
    \resizebox*{\textheight}{!}{
	\begin{tabular}{@{} ccccccccccc @{}}
		\hline
		label position & label & additive & a & c$_{ab}$ & c$_{b1}$ & c$_{b2}$ & $\tau_{ab}$  & $\tau_{b1}$ & $\tau_{b2}$ &  $\tau_0$\\
		&  &  &  &  &  &  & / ns & / ns & / ns & / ns \\

		\hline \hline
		61 & Atto550 & apo  & 114.0$\pm$0.6 & 0.82$\pm$0.08 & 0.10$\pm$0.011 & - & 1.8$\pm$0.3  & 112$\pm$27 & - & -0.6$\pm$0.11\\

		61 & Atto550 & AMPPNP & 26.9$\pm$0.2 & 0.75$\pm$0.06 & 0.106$\pm$0.008 & - & 1.9$\pm$0.2  & 164$\pm$36 & - & -0.7$\pm$0.10\\

		298 & Atto532 & apo & 71.88$\pm$0.07 & 0.8$\pm$1 & 0.5$\pm$7 & - & 3.00$\pm$0.02  & 5$\pm$12 & - & -1.16$\pm$0.06\\

		298 & Atto532 & AMPPNP & 22.68$\pm$0.02 & 0.883$\pm$0.007 & 2.7$\pm$0.14 & 0.03$\pm$0.011 & 8.3$\pm$0.6  & 5.2$\pm$0.16 & 65$\pm$22  & -0.7$\pm$0.03\\

		298 & Atto532 & Sba1, AMPPNP & 5.238$\pm$0.005  & 0.76$\pm$0.012  & 1.9$\pm$0.13 & 0.037$\pm$0.004 & 7.0$\pm$0.6  & 5.2$\pm$0.2 & 100$\pm$18 & -0.66$\pm$0.04\\

		385 & Atto532 & apo & 22.65$\pm$0.06 & 0.7$\pm$3 & 0.5$\pm$14 & 0.04$\pm$0.011 & 2.75$\pm$0.03  & 5$\pm$24 & 101$\pm$46 & -0.65$\pm$0.08\\

		385 & Atto532 & AMPPNP & 20.4$\pm$0.12 & 0.85$\pm$0.010 & 2.4$\pm$0.2 & 0.029$\pm$0.004 & 7.5$\pm$0.7 & 5.3$\pm$0.2 & 246$\pm$133 & -0.70$\pm$0.05\\

		452 & Atto550 & apo & 47.6$\pm$0.3 & 0.83$\pm$0.06 & 0.104$\pm$0.009 & - & 2.2$\pm$0.2 & 133$\pm$29 & - & -0.7$\pm$0.10\\

		452 & Atto550 & AMPPNP & 49.7$\pm$0.4 & 0.76$\pm$0.04 & 0.097$\pm$0.007 & - & 2.0$\pm$0.2 & 204$\pm$43 & - & -0.66$\pm$0.08\\
		\hline
	\end{tabular}}
	\label{SI:tab:nsFCShsp90donoronly}
\end{sidewaystable}

\begin{sidewaystable}[p!]
	\subsection*{Summary of nsFCS fit results for free dyes and tryptophan}
	\centering
	\caption{\baselineskip4mm Summary of nsFCS results on freely diffusing Atto532, Atto550 and Atto647N with and without tryptophan. `Correlation' specifies the detection channels which were correlated to obtain the correlation function G($\tau$). DxD and AxA stand for the cross-correlations of the parallel and perpendicular donor and acceptor channels, respectively. Cross-correlation of different polarization channels removed artefacts resulting from detector afterpulsing. The used fit models are indicated by capital letters (A, B, C) and differ by the number of bunching components used. Fit parameters are a scaling factor $a$, the weight of the antibunching mode $c_{ab}$, the weights of the bunching modes $c_{b1}$ and $c_{b2}$, antibunching time $\tau_{ab}$, bunching times $\tau_{b1}$ and $\tau_{b2}$, respectively. $\tau_0$ corrects for a small time delay between the detection channels. Data were analysed on the fit interval from 0 to 100\,ns.}
	\begin{tabular}{@{} ccccccccc @{}}
		\hline
		fluorophore & Trp  & correlation  & a & c$_{ab}$ & c$_{b}$ & $\tau_{ab}$  & $\tau_{b}$  &  $\tau_0$\\
		& / mM &              &   &          &         & / ns         & / ns        & / ns \\

		\hline \hline

		Atto532 & 0   & DxD & 13.00$\pm$0.013 & 0.58$\pm$0.015 &   & 2.00$\pm$0.07  &  &   -0.57$\pm$0.03\\

		& 10 & DxD & 40.59$\pm$0.07 & 0.84$\pm$0.02 & 1.6$\pm$0.4  & 6$\pm$1  & 5.5$\pm$0.4  & -0.57$\pm$0.04\\

		& 20  & DxD & 82.1$\pm$0.2 & 0.87$\pm$0.02 & 2.0$\pm$0.5    & 6$\pm$1    & 5.6$\pm$0.3  & -0.60$\pm$0.05\\

		& 30  & DxD & 47.6$\pm$0.13 & 0.87$\pm$0.02 & 2.3$\pm$0.6     & 6$\pm$2  & 5.7$\pm$0.3   & -0.62$\pm$0.05\\

		& 40 & DxD & 81.9$\pm$0.2 & 0.86$\pm$0.05 &  2$\pm$1  & 5$\pm$2  & 5.5$\pm$0.2 &    -0.50$\pm$0.05\\

		& 50 & DxD & 59.9$\pm$0.3 & 0.87$\pm$0.03 &  2.9$\pm$0.9   & 6$\pm$2  &  4.9$\pm$0.3   & -0.62$\pm$0.08\\

		& 60 & DxD & 110.0$\pm$0.4 & 0.88$\pm$0.04 &   3$\pm$1   & 5$\pm$2  & 5.1$\pm$0.2    & -0.57$\pm$0.06\\

		\hline

		Atto550 & 0   & DxD & 244.6$\pm$0.6 & 0.93$\pm$0.04 &   & 2.2$\pm$0.12  &  &   -0.59$\pm$0.06\\

		& 10 & DxD & 912$\pm$7 & 0.96$\pm$0.05 & 0.4$\pm$0.10  & 2.5$\pm$0.4  & 15$\pm$4  & -0.68$\pm$0.09\\

		& 20  & DxD & 264.3$\pm$0.3 & 0.91$\pm$0.03 & 0.71$\pm$0.08    & 2.5$\pm$0.3    & 13$\pm$1  & -0.61$\pm$0.05\\

		& 30  & DxD & 64.5$\pm$0.3 & 0.67$\pm$0.03 & 0.7$\pm$0.12     & 2.6$\pm$0.5  & 11$\pm$1   & -1.14$\pm$0.09\\

		& 40 & DxD & 96.9$\pm$0.6 & 0.83$\pm$0.04 &  3$\pm$1  & 5$\pm$2  & 8$\pm$3 &    -1.1$\pm$0.12\\
		\hline

		Atto647N   & 0   & AxA & 49.12$\pm$0.06 & 0.83$\pm$0.014 & - & 3.23$\pm$0.08  & - &   0.09$\pm$0.04\\

		& 10 & AxA & 200.1$\pm$0.5 & 0.91$\pm$0.03 & - & 3.1$\pm$0.15  & - &    0.21$\pm$0.07\\

		& 20  & AxA & 1767$\pm$11 & 0.90$\pm$0.07 & - & 2.6$\pm$0.3  & - &    -0.03$\pm$0.16\\

		& 30  & AxA & 1185$\pm$7 & 0.96$\pm$0.07 & - & 2.8$\pm$0.3  & - &    0.2$\pm$0.1\\

		& 40 & AxA & 66.7$\pm$0.14 & 0.79$\pm$0.02 & - & 2.9$\pm$0.12  & - &    0.13$\pm$0.06\\

		& 65 & AxA & 50.0$\pm$0.11 & 0.74$\pm$0.03 & - & 2.9$\pm$0.14  & - &    0.09$\pm$0.07\\
		\hline
	\end{tabular}
	\label{SI:tab:nsFCSdyes}
\end{sidewaystable}

\begin{table}[!h]
	\subsection*{Fit results of single-molecule time-resolved anisotropy analysis 1}
	\centering
	\caption{\baselineskip 4mm Acceptor anisotropy analysis at Hsp90 position 452. The Hsp90 FRET pair 452-452 was labelled with Atto550 and Atto647N and measured with AMPPNP at 21$^{\circ}$C. FRET vs. stoichiometry analysis was used to assign conformational sub-states of Hsp90 (open, closed\,A, closed\,B). Data were analysed using the \textit{cone-in-cone} model \autocite{Schroder.2005}:  $\rho_{\textrm{dye}}$ describes free dye rotation, $\rho_{\textrm{local}}$ the rotation of structural elements to which Atto647N was attached to (e.g. a loop) and $\rho_{\textrm{global}}$ the global rotation of the overall protein. }
	\begin{tabular}{@{} cccccccc @{}}
		\hline
		state & $r_0$ & $A_{\textrm{dye}}$ & $A_{\textrm{local}}$ & $\rho_{\textrm{dye}}$ & $\rho_{\textrm{local}}$  & $\rho_{\textrm{global}}$ & R$^2$\\
		&       &       &       & / ns     & / ns      & / ns & \\
		\hline \hline
		open  &  0.80 & 0.43$\pm$0.22 & 0.66$\pm$0.04 & 0.25$\pm$0.12 & 1.8$\pm$0.3 & 66$\pm$6 & 0.93  \\
		A  & 0.68 & 0.40$\pm$0.04 & 0.70$\pm$0.03 & 0.39$\pm$0.07 & 2.7$\pm$0.7 & 77$\pm$14 & 0.96 \\

		B & 0.49 & 0.51$\pm$0.08 & 0.92$\pm$0.17 & 0.61$\pm$0.20 & 3$\pm$6 & 116$\pm$41  & 0.81\\
		\hline
	\end{tabular}
	\label{SI:tab:aniso}
\end{table}

\begin{table}[!h]
		\subsection*{Fit results of single-molecule time-resolved anisotropy analysis 2}
	\centering
	\caption{\baselineskip 4mm Donor anisotropy analysis of Hsp90 and AMPPNP in presence and absence of Sba1. Hsp90 was labeled with Atto532 at position 298 and measured at 21$^{\circ}$C. Data were analysed using the \textit{cone-in-cone} model \autocite{Schroder.2005}:  $\rho_{\textrm{dye}}$ describes free dye rotation, $\rho_{\textrm{local}}$ the rotation of structural elements to which Atto532 was attached to (e.g. a loop) and $\rho_{\textrm{global}}$ the global rotation of the overall protein complex. }
	\begin{tabular}{@{} cccccccc @{}}
		\hline
		additive & $r_0$ & $A_{\textrm{dye}}$ & $A_{\textrm{local}}$ & $\rho_{\textrm{dye}}$ & $\rho_{\textrm{local}}$  & $\rho_{\textrm{global}}$ & R$^2$\\
		&       &       &       & / ns     & / ns      & / ns & \\
		\hline \hline
		AMPPNP  &  0.47 & 0.62$\pm$0.03 & 0.74$\pm$0.04 & 0.40$\pm$0.06 & 1.5$\pm$0.3 & 200$\pm$43& 0.99  \\
		AMPPNP, Sba1  &  0.47 & 0.70$\pm$0.01 & 0.73$\pm$0.01 & 0.45$\pm$0.03 & 2.2$\pm$0.2 & 200$\pm$32& 0.99  \\
		\hline
	\end{tabular}
	\label{SI:tab:anisoHsp90Sba1}
\end{table}

\begin{table}[!h]
	\subsection*{Fit results of MD-based accessible dye volume auto-correlations}
	\centering
	\caption{\baselineskip4mm Fit results of MD-based accessible dye volume correlations analysis. Accessible dye volumes were auto-correlated for different Hsp90 positions based on 1\,µs-MD traces with AMPPNP (A = chain A, B = chain B).}
	\begin{tabular}{@{} cccccc @{}}
		\hline
		correlation & $A_1$ & $A_2$ & $\tau_1$ &  $\tau_2$ & R$^2$\\

		        	&       &       & / ns     & / ns      &  \\
		\hline \hline
		A61 x A61	& $(69\pm6)\cdot10^{9}$ & $(66.8\pm0.4)\cdot10^{9}$ & $3\cdot10^{-5}\pm$NaN & $103\pm1$ & $0.97$ \\

		B61 x B61	& $(120\pm7)\cdot10^{9}$ & $(70.0\pm0.6)\cdot10^{9}$ & $0.16\pm$NaN & $72\pm1$ & $0.95$ \\

		A298 x A298	& $(120\pm4)\cdot10^{9}$ & $(70.0\pm0.7)\cdot10^{9}$ & $1.0\pm$NaN & $80\pm1$ & $0.95$ \\

		A452 x A452	& $(120\pm10)\cdot10^{9}$ & $(49.5\pm0.7)\cdot10^{9}$ & $3\cdot10^{-5}\pm$NaN & $100\pm2$ & $0.78$ \\

		B452 x B452	& $(120\pm5)\cdot10^{9}$ & $(70.0\pm0.8)\cdot10^{9}$ & $0.9\pm$NaN & $75\pm1$ & $0.92$ \\
		\hline
	\end{tabular}
	\label{SI:tab:AVcorr}
\end{table}